\documentclass[fleqn,usenatbib]{mnras}

\usepackage{newtxtext,newtxmath}

\usepackage[T1]{fontenc}
\usepackage{soul}
\usepackage{mathtools}

\DeclareRobustCommand{\VAN}[3]{#2}
\let\VANthebibliography\thebibliography
\def\thebibliography{\DeclareRobustCommand{\VAN}[3]{##3}\VANthebibliography}

\usepackage{graphicx}	
\usepackage{amsmath}	

\title[]{Dynamics or Geysers and tracer transport over the south pole of Enceladus}

\author[Kang et al.]{
  Wanying Kang,$^{1}$\thanks{E-mail: wanying@mit.edu}
  John Marshall,$^{1}$
  Tushar Mittal,$^{1}$
  and Suyash Bire$^{1}$
\\
$^{1}$Earth, Atmospheric and Planetary Science Department, Massachusetts Institute of Technology, Cambridge, MA 02139, USA
}

\pubyear{2022}

\begin{document}
\label{firstpage}
\pagerange{\pageref{firstpage}--\pageref{lastpage}}
\maketitle

\begin{abstract}
  Over the south pole of Enceladus, an icy moon of Saturn, geysers eject water into space in a striped pattern, making Enceladus one of the most attractive destinations in the search for extraterrestrial life. We explore the ocean dynamics and tracer/heat transport associated with geysers as a function of the assumed salinity of the ocean and various core-shell heat partitions and bottom heating patterns. We find that, even if heating is concentrated into a narrow band on the seafloor directly beneath the south pole, the warm fluid becomes quickly mixed with its surroundings due to baroclinic instability. The warming signal beneath the ice is diffuse and insufficient to prevent the geyser from freezing over. Instead, if heating is assumed to be local to the geyser, emanating from tidal dissipation in the ice itself, the geyser can be sustained. In this case, the upper ocean beneath the ice becomes stably stratified and thus a barrier to vertical communication, leading to transit timescales from the core to the ice shell of hundreds of years.
\end{abstract}

\begin{keywords}
planets and satellites: oceans, planets and satellites: interiors
\end{keywords}


\section{Introduction}
\label{sec:introduction}
Despite its small size (only 252 km in radius with a surface area roughly the same as Texas) and hence rapid heat loss, Enceladus retains an approximately 40 km deep global ocean beneath its ice shell \citep{Thomas-Tajeddine-Tiscareno-et-al-2016:enceladus}. Geyser-like jets are ejected into space over the south polar region \citep{Porco-Helfenstein-Thomas-et-al-2006:cassini, Howett-Spencer-Pearl-et-al-2011:high, Spencer-Howett-Verbiscer-et-al-2013:enceladus, Iess-Stevenson-Parisi-et-al-2014:gravity}, providing a unique opportunity to peek through the $\sim 20km$-thick ice shell \citep{Beuthe-Rivoldini-Trinh-2016:enceladuss, Hemingway-Mittal-2019:enceladuss} into the ocean below. Several flybys accomplished by the Cassini mission have yielded vast amounts of data, greatly improving our understanding of this small but active icy satellite. Particles and gases sampled from these jets indicate the presence of organic matter \citep{Postberg-Khawaja-Abel-et-al-2018:macromolecular}, silica nanoparticles \citep{Hsu-Postberg-Sekine-et-al-2015:ongoing} and a modestly alkaline environment \citep{Glein-Baross-Waite-2015:ph}, all suggestive of seafloor hydrothermal activity and astrobiological potential \citep{Glein-Postberg-Vance-2018:geochemistry, Taubner-Pappenreiter-Zwicker-et-al-2018:biological, McKay-Davila-Glein-et-al-2018:enceladus}. In the planning of future missions \citep{Tsou-Brownlee-McKay-et-al-2012:life, Mora-Jones-Creamer-et-al-2018:extraction, MacKenzie-Neveu-Davila-et-al-2021:enceladus}, it is crucial to explore
whether any nutrients or biosignatures can be transported by the ocean from the core to geyser regions, and on what timescales, and how to make best use of such information to infer the nature of the geochemical environment of the subsurface ocean. To answer these questions, we need to know not only the physical/chemical state of the ocean, but also the nature of dynamical controls on the efficiency of tracer transport.

The ocean on icy moons is forced by salinity and heat fluxes at the ice-ocean boundary and heating at the seafloor. Just under the ice shell, the ocean temperature is close to the local freezing point of water, which is higher near the geysers where the ice is thin and the pressure is lower (see Eq.~\eqref{eq:freezing-point}). In addition, the freezing/melting of ice will change the local salinity through brine rejection/freshwater production. There will thus be temperature and salinity gradients at the upper surface which will in turn induce ocean circulation. At the ocean bottom, hydrothermal activity powered by tidal heating in the silicate core will warm up the ocean. The magnitude of the density and hence pressure anomalies that ultimately drive circulation depend on both temperature and salinity gradients, and this dependency varies with salinity: water expands upon warming in a salty ocean but contracts upon warming in a fresh ocean \citep{Kang-Mittal-Bire-et-al-2021:how}. When the latter occurs (anomalous expansion), convection will not ensue until a critical temperature ($\sim$2K in excess of the freezing point on Enceladus) is reached. What is more, the partition of heating between the core and the ice shell determines the ocean stratification and whether convection can be triggered.

Currently, both salinity and core-shell heat partition remain poorly constrained. Observations show that both the volume of fluid emitted by the plumes and the thermal emission of the moon have strong diurnal cycles, suggesting a marked tidal modulation \citep{Hedman-Gosmeyer-Nicholson-et-al-2013:observed, Nimmo-Porco-Mitchell-2014:tidally, Porco-DiNino-Nimmo-2014:how, Teolis-Perry-Hansen-et-al-2017:enceladus, Hansen-Esposito-Aye-et-al-2017:investigation, Hurford-Helfenstein-Hoppa-et-al-2007:eruptions}. Numerical studies have shown that the tidal heating in the ice over the south pole can be comparable to the observed heat flux due to the combined action of faults and ice thinning \citep{Soucek-Behounkova-Cadek-et-al-2019:tidal}. However, heating in the ice alone may not be enough to prevent the ocean from freezing given our current understandings in ice rheology \citep{Beuthe-2019:enceladuss}. Use of advanced rheology models can result in greater dissipation rates, but may still be insufficient to account for global heat loss rates \citep{McCarthy-Cooper-2016:tidal,Renaud-Henning-2018:increased}. On the other hand, there are model calculations suggesting that strong heating may be occurring in the silicate core \citep{Choblet-Tobie-Sotin-et-al-2017:powering, Liao-Nimmo-Neufeld-2020:heat}. If core heating indeed dominates, one would expect the seafloor to be hydrothermally active, which in turn may provide the chemical gradients required to support life \citep{Deamer-Damer-2017:can}. That said, the uncertainty associated with the core rheology is even higher, and if tidal heating in the ice is weak, such that most of the heat is produced in the core, equatorial heat transport by ocean circulation is likely to flatten out ice thickness variations via the ice pump mechanism \citep{Kang-Mittal-Bire-et-al-2021:how}. In the absence of direct measurements of tidal dissipation rates in the core and shell, it seems useful to ask what core-shell heat partitioning and bottom heating patterns might be required to sustain geysers. This is one of the main goals of the present study.

Early estimates of the salinity of Enceladus' ocean are based on assumptions of thermochemical equilibrium. Considering a range of hydrothermal and freezing conditions for chondritic compositions, a salinity between $2$-$20$~psu (g/kg) is implied \citep{Zolotov-2007:oceanic, zolotov2014can, Glein-Postberg-Vance-2018:geochemistry}. However, at least $\sim 17$ psu is required to keep the geysers' liquid–gas interface convectively active ensuring that they do not freeze up \citep{Ingersoll-Nakajima-2016:controlled}. Sodium-enriched samples taken by Cassini from south pole sprays have a salinity of 5-20~psu. This can be considered a lower bound since the interaction of cold water vapor sprays with their environment may lower the salinity of droplets through condensation \citep{Postberg-Kempf-Schmidt-et-al-2009:sodium}. There are uncertainties, however, since fractional crystallization and disequilibrium chemistry may partition components in such a way that geyser particles are not directly representative of the underlying ocean \citep{Fox-Powell-Cousins-2021:partitioning}. Furthermore, if particles originate from a hydrothermal vent, composition can deviate far from that of the overall ocean \citep{Glein-Postberg-Vance-2018:geochemistry, Choblet-Tobie-Sotin-et-al-2017:powering}. Remarkably, the size of silica nano-particles carried along in the sprays can also be used to estimate ocean salinity. Assuming an intermediate value of pH and short transport timescale via hydrothermal vents, a salinity $<40$~psu is obtained \citep{Hsu-Postberg-Sekine-et-al-2015:ongoing}. In a separate line of argument, oceans with too much or too little salt may have a strong ice pump effect, leading the erosion of ice thickness gradients \citep{Kang-Mittal-Bire-et-al-2021:how}.

In this study, we will explore the plume dynamics and tracer transport under various ocean salinities, core-shell heat partitions and bottom heating patterns and investigate 1) scenarios that can keep the geyser from freezing up, and 2) how long it takes for tracers to travel from the seafloor upward to the geyser regions.

\section{Methods}
\label{sec:methods}
To study the small-scale ocean dynamics and transport timescales associated with geysers on Enceladus, we use the state-of-the-art Massachusetts Institute of Technology OGCM (MITgcm) \citep{MITgcm-group-2010:mitgcm, Marshall-Adcroft-Hill-et-al-1997:finite} to simulate the 3D ocean dynamics near one geyser stripe at 100-m resolution. The same model and algorithm (ref to Oceananigans) has been used to study convection in the ocean (refs) and convection on icy moons (refs). In an advance on previous studies \citep{Soderlund-2019:ocean}, here we account for the two-way coupling between the ice shell and ocean by allowing heat exchange between ocean and ice, resulting in freezing/melting of ice with its concomitant heat/salt flux. This, in turn, drives ocean dynamics. The geyser stripe is prescribed at the center of the domain along $y=0$ ($-L_y/2\leq y\leq Ly/2$, $L_y=32$~km) as a Gaussian-shaped indentation on the underside of the ice shell with no variation in $x$. This ice geometry does not evolve to any significant extent on the timescales considered here. Because the ice shell is thinner near the geyser, heat is conducted through the ice more easily (see Eq.~\ref{eq:heat-condution}). However, heating due to dissipation in the ice is also stronger due to the rheology feedback, which partially offsets heat conduction \citep{Beuthe-2019:enceladuss} (see Eq.~\ref{eq:H-tide}). In addition, the seawater near the geyser will be warmer because the freezing point rises with decreasing pressure (see Eq.~\ref{eq:freezing-point}). These features distinguish the geyser regions from elsewhere, driving motions and heat/salt exchanges between the two. To facilitate integration out to steady state, a two-step approach is taken. We first integrate an equivalent 2D configuration ($y\times z$) for several hundred years. This is then used to initialise the 3D model which is integrated on for another 200 years or so.

Due to the uncertainties associated with the ocean mean salinity \citep{Zolotov-2007:oceanic, zolotov2014can, Glein-Postberg-Vance-2018:geochemistry, Ingersoll-Nakajima-2016:controlled, Postberg-Kempf-Schmidt-et-al-2009:sodium, Fox-Powell-Cousins-2021:partitioning, Glein-Postberg-Vance-2018:geochemistry, Choblet-Tobie-Sotin-et-al-2017:powering, Kang-Mittal-Bire-et-al-2021:how}, the core-shell heat partitioning \citep{Travis-Schubert-2015:keeping, Choblet-Tobie-Sotin-et-al-2017:powering, Hemingway-Mittal-2019:enceladuss, Kang-Mittal-Bire-et-al-2021:how}, and the heating pattern at the seafloor \citep{Choblet-Tobie-Sotin-et-al-2017:powering}, six different scenarios are explored. In the first four, a uniform bottom heating pattern is assumed, and we explore combinations of two ocean salinities ($10,\ 30$~psu) and two core-shell heat partitions ($10,\ 90$\% heating in the core). Hereafter, we refer to them as S$x$c$y$, where $x$ is the salinity and $y$ is the percentage of heating in the core. Heating in the ice will naturally be concentrated near the poles due to the rheology feedback (see Eq.~\ref{eq:H-tide}) \citep{Beuthe-2019:enceladuss, Kang-Flierl-2020:spontaneous}, while the core heating could be both distributed (due to the high porosity) or focused \citep{Choblet-Tobie-Sotin-et-al-2017:powering}. When the bottom heating is strong ($c=90$\%), we expect its assumed pattern to be of importance. Therefore two additional scenarios are considered: S30c90v and S10c90v, where ``v'' indicates that the heating is concentrated in a narrow vent. In these two experiments, bottom heating is confined to a narrow band (not unlike the distribution hypothesised by Choblet et al) situated directly underneath the geyser, with a peak heat flux of 10~W/m$^2$. This is likely the strongest concentrated flux which can be achieved by hydrothermal systems on Enceladus \citep{Choblet-Tobie-Sotin-et-al-2017:powering}. Experiments S30c90v and S10c90v enable us to test whether or not a bottom-concentrated heating can be transmitted to the ice without being mixed away, making it more likely that the geyser can be kept open. Snapshots of the dynamic and thermodynamic state for each scenario are presented in the appendix.

\section{Results}
\label{sec:results}

\subsection{Conditions required to sustain the geyser}
\label{sec:geyser-sustainability}

\begin{figure*}
    \centering
    \includegraphics[page=9,width=\textwidth]{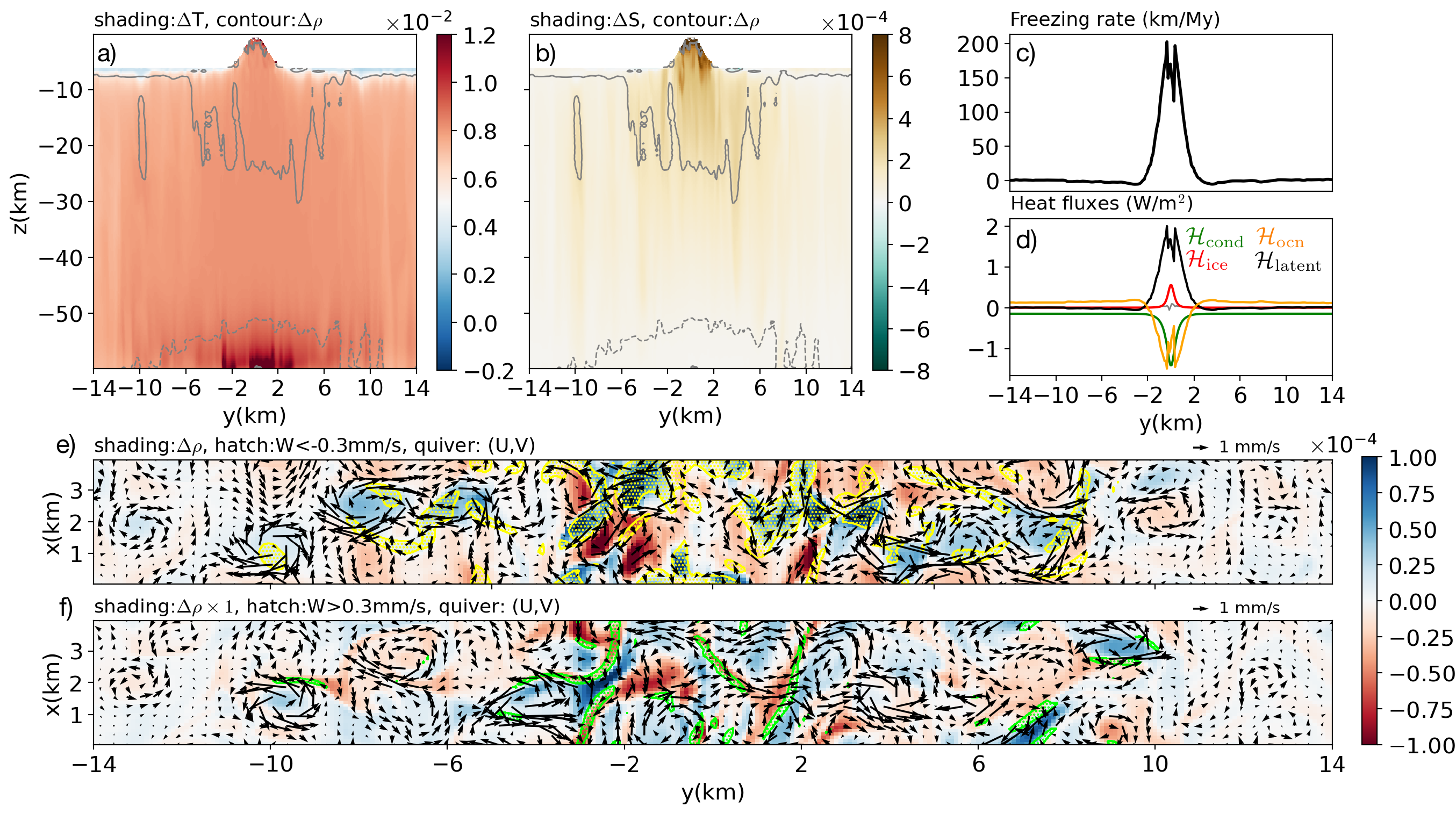}
    
    \caption{\small{Snapshot taken at the end of the simulation for the S30c90v scenario. Panel (a) shows the temperature anomalies at a given $x$ in shading and density anomalies in contours. Solid contour denote positive density anomaly and dashed contours denote negative density anomaly. From thin to thick, contours mark $\Delta \rho=\pm 10^{-4}$, $\pm 8\times 10^{-4}$~kg/m$^3$, $\pm 5\times 10^{-3}$~kg/m$^3$, $\pm 5\times 10^{-2}$~kg/m$^3$. Panel (b) is similar to panel (a) except salinity is shown in place of temperature. Panel (c,d) shows the freezing/melting rate and heat budget of the ice shell, respectively. In panel (d), red, orange, green and black curves, respectively, represent the ice dissipation $\mathcal{H}_{\mathrm{ice}}$, the heat absorbed from the ocean $\mathcal{H}_{\mathrm{ocn}}$, the conductive heat loss through the ice $\mathcal{H}_{\mathrm{cond}}$ and the latent heat release $\mathcal{H}_{\mathrm{latent}}$. The gray dashed curve shows the residue of the heat budget, which should be close to zero. Panel (e,f) show the dynamics in a horizontal plane, horizontal flow speeds in quivers, density anomaly in the shading, and areas with vertical speed beyond a certain threshold (see text just above the figure) are marked by hatches. Green hatches denote upward motions and yellow hatches denote downward motions. The plane shown by panel (e) is taken just below the ice shell (z=-9km), and the plane shown by panel (f) is just above the seafloor (z=-56km).}}
    
    \label{fig:S30c90v-dynamics}
  \end{figure*}
  
In order to keep the geyser open, heat needs to be concentrated toward it to compensate for strong conductive heat loss (Eq.~\ref{eq:H-cond}). This is true irrespective of whether the heat source is in the ice shell or the core. If the core is the primary heat source, the ocean must carry that heat to the ice shell while keeping it concentrated. In a salty ocean, bottom heating increases the buoyancy of water ($\alpha>0$), which may trigger convective plumes that penetrate the entire ocean depth and deliver heat to the geysers. To test this scenario, we prescribe the bottom heating to be perfectly aligned with the geyser atop. As can be seen from Fig.\ref{fig:S30c90v-dynamics}, rather than remaining concentrated, the convective plumes become turbulent shortly after departing from the seafloor. By the time they have reached upward 10km or so, turbulence begins to fill the entire domain as lateral instabilities spread the temperature signal horizontally.

The turbulence stems from baroclinic instability along the front that separates the warm plume water from the cold environment. Convection is always susceptible to baroclinic instability when the buoyancy source that triggers it is localised in space, as here. The phenomenon has been well documented in laboratory and theoretical studies \citep{Saunders-1973:instability, Jones-Marshall-1993:convection, Maxworthy-Narimousa-1994:unsteady, Visbeck-Marshall-Jones-1996:dynamics, Legg-Jones-Visbeck-1996:heton, Jacobs-Ivey-1998:influence, Jacobs-Ivey-1999:rossby, Bush-Woods-1999:vortex, Okada-Ikeda-Minobe-2004:numerical}. It has been foud that, when the domain is deep enough for instability to take place, a line plume tends to break into a chain of vortices with a diameter of
\begin{equation}
    \label{eq:deformation-radius}
    L_f\sim 11(B_L/f^3)^{1/3},
  \end{equation}
  where $B_L=\int_{-L_y/2}^{L_y/2} B_A~dy$ is the buoyancy flux per length \citep{Bush-Woods-1999:vortex}, $B_A=\alpha g Q_A/(\rho C_p)$ is the buoyancy flux per area, $g$ is gravity, $\alpha=4\times10^{-5}$/K is the thermal expansion coefficient near the bottom, $Q_A$ is the heat flux per area, and $f$ is the Coriolis parameter. These criteria guarantee that the circumference of the plume is sufficiently large, and the travel time across the domain depth sufficiently long, for instability to grow. Substituting parameters from our simulation yields a vortex diameter of $L_f\sim 200$~m, suggesting an instability if the plume length is greater than 400~m. This condition is well satisfied since the seafloor vent is assumed to have the same shape as the ``tiger stripe'' seen at the surface, which is hundreds of kilometers in length. In the presence of such an instability, the heat flux will be spread out and diluted before it can affect the ice shell. As seen in Fig.~\ref{fig:S30c90v-dynamics}f, the plume indeed splits up into vortices and turbulence. As a result, the heated water is spread over a much wider area than that of the initial source, which has a Gaussian distribution with a standard deviation of $\sigma_{\mathrm{core}}=200$~m (Fig.~\ref{fig:S30c90v-dynamics}a). The diameter of the vortices observed in the simulation is around 1-2~km, which is 5 times greater than that predicted by the above scaling law. This difference may be due to the relatively large viscosity (0.05~m$^2$/s) employed which suppresses the instability on smaller scales. The vortices obtained are just small enough to fit into the 4~km wide domain. If the domain width is cut by half, only one wavelength is allowed and the flow remain laminar (see Fig.~\ref{fig:S30c90v-narrow}a), and when we double the domain size to the default, instability shows up (Fig.~\ref{fig:S30c90v-narrow}b).

\begin{figure*}
    \centering
    \includegraphics[page=10,width=\textwidth]{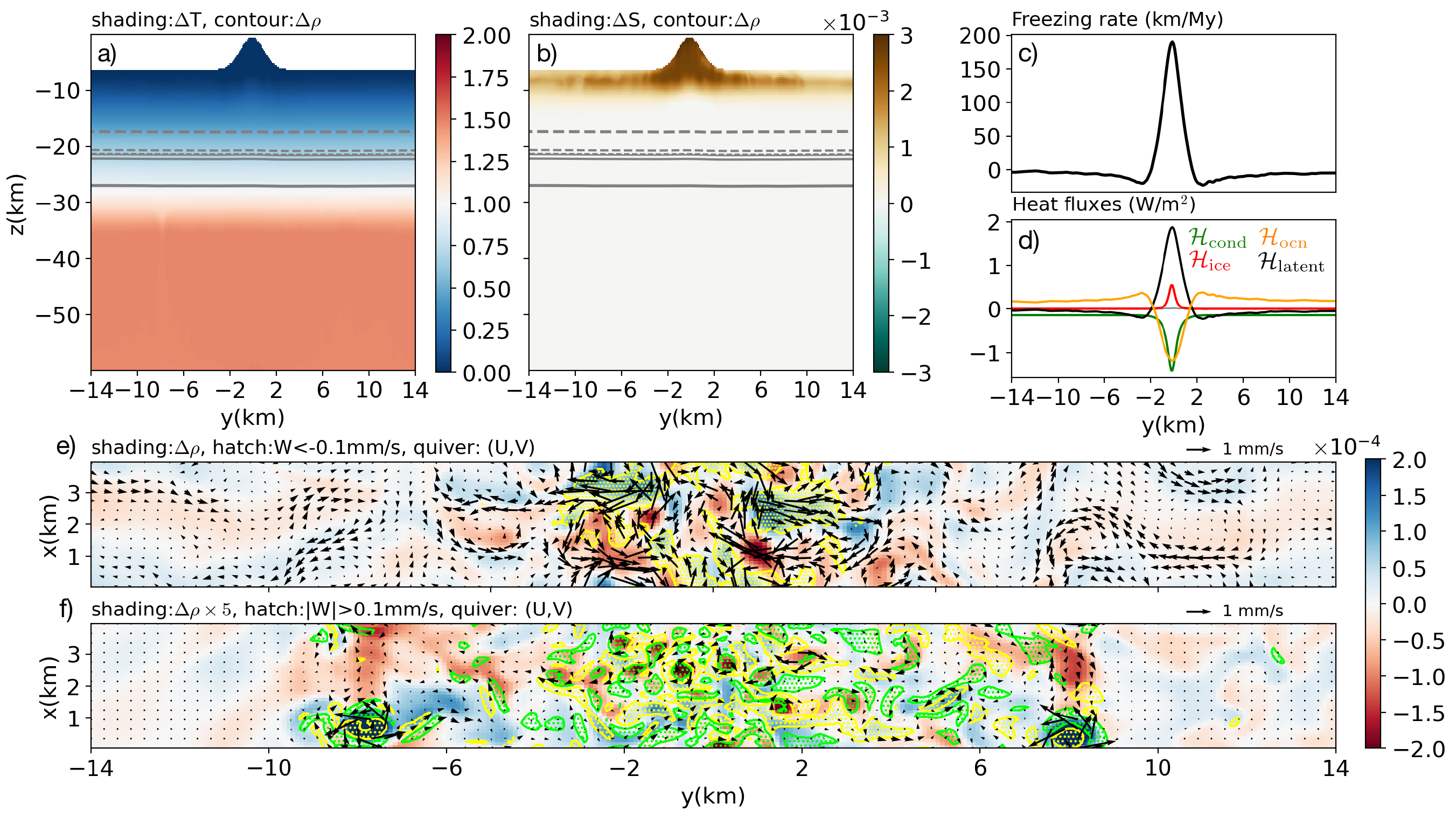}
    
    \caption{\small{Same as Fig.~\ref{fig:S30c90v-dynamics} except for S10c90v.}}
    
    \label{fig:S10c90v-dynamics}
  \end{figure*}

In a fresh ocean ($S_0<22$~psu), convection is suppressed and warmed fluid is confined to the bottom. This is because fresh water contracts upon warming when it is near the freezing point at low pressure: bottom heating stabilizes rather than destabilizes the water column. In the presence of this stably stratified layer (see Fig.~\ref{fig:S10c90v-dynamics}a), any concentration of heat flux is again likely to be homogenized before reaching the ice shell. It should be noted that in the lower part of the ocean, where high temperatures and pressures suppress anomalous expansion, convective plumes are present and shoot upwards before breaking up, just as seen in  experiment S30c90v (Fig.~\ref{fig:S10c90v-dynamics}f). However, none of these plumes can penetrate into the strongly stratified layer above, where heat is uniformly diffused upward. Due to the low ocean salinity, the thermal expansion coefficient near the seafloor is 1-2 orders of magnitude smaller than the high salinity scenario and decays further with height. As a result, $R_d$ is perhaps 2-3 times smaller, as reflected in the rather small size of the plumes observed in the experiment.

Thus, whether the water is salty or fresh, we expect heat to be rather evenly delivered to the ice shell even if the source is concentrated at the bottom. As a consequence, the region of the geyser inevitably freezes because heat loss there occurs at a greater rate than elsewhere (Fig.~\ref{fig:S30c90v-dynamics}c,~\ref{fig:S10c90v-dynamics}c). Through brine rejection, freezing makes the water salty and thus dense (Fig.~\ref{fig:S30c90v-dynamics}b,~\ref{fig:S10c90v-dynamics}b), triggering convection from the top. The associated buoyancy flux per length $B_L$ can be estimated thus: $\int_{-\sigma_H}^{\sigma_H}\beta gqS_0~dy \approx 2 \sigma_H \beta g \bar{q} S_0\approx 2\times 10^{-8}$~m$^3$/s$^3$. This is comparable to the buoyancy flux induced by hydrothermal heating at the bottom. Here $q$ is the freezing rate, $\bar{q}\sim 100$~km/My near the geyser, $\sigma_H=1$~km is the width of the geyser and $\beta$ is the haline contraction coefficient at mean salinity $S_0$. This explains why the size of the salinity-driven vortex near the top is similar to that of the heat-driven vortex near the bottom --- Fig.~\ref{fig:S30c90v-dynamics}h. As the downward convective plume breaks up, turbulence is generated, spreading the salty water into the surroundings (Fig.~\ref{fig:S30c90v-dynamics}b), while mixing the warm water near the geyser with the cold water under the thicker ice shell on either side (see Fig.~\ref{fig:S30c90v-dynamics}a and Fig.~\ref{fig:S10c90v-dynamics}a). As a result, heat is transported away from the geyser area, as quantified by the ice shell heat budget (see orange curve in Fig.~\ref{fig:S30c90v-dynamics}d and Fig.~\ref{fig:S10c90v-dynamics}d). This almost doubles the geyser's freezing rate triggered by the initial conductive heat loss (green curve). 

To summarize, regardless of ocean salinity, concentrated heating from the bottom is unlikely to remain concentrated when transmitted to the ice shell, and inevitably leads to freezing of the geyser. When the bottom heating is not localized (S30c90 and S10c90), the freezing rate and heat budgets are remarkably similar to that obtained when it is concentrated (S30c90v and S10c90v), indicating that the bottom heating pattern is ``forgotten'' as it is transmitted upward.

\begin{figure*}
    \centering
    \includegraphics[page=11,width=0.8\textwidth]{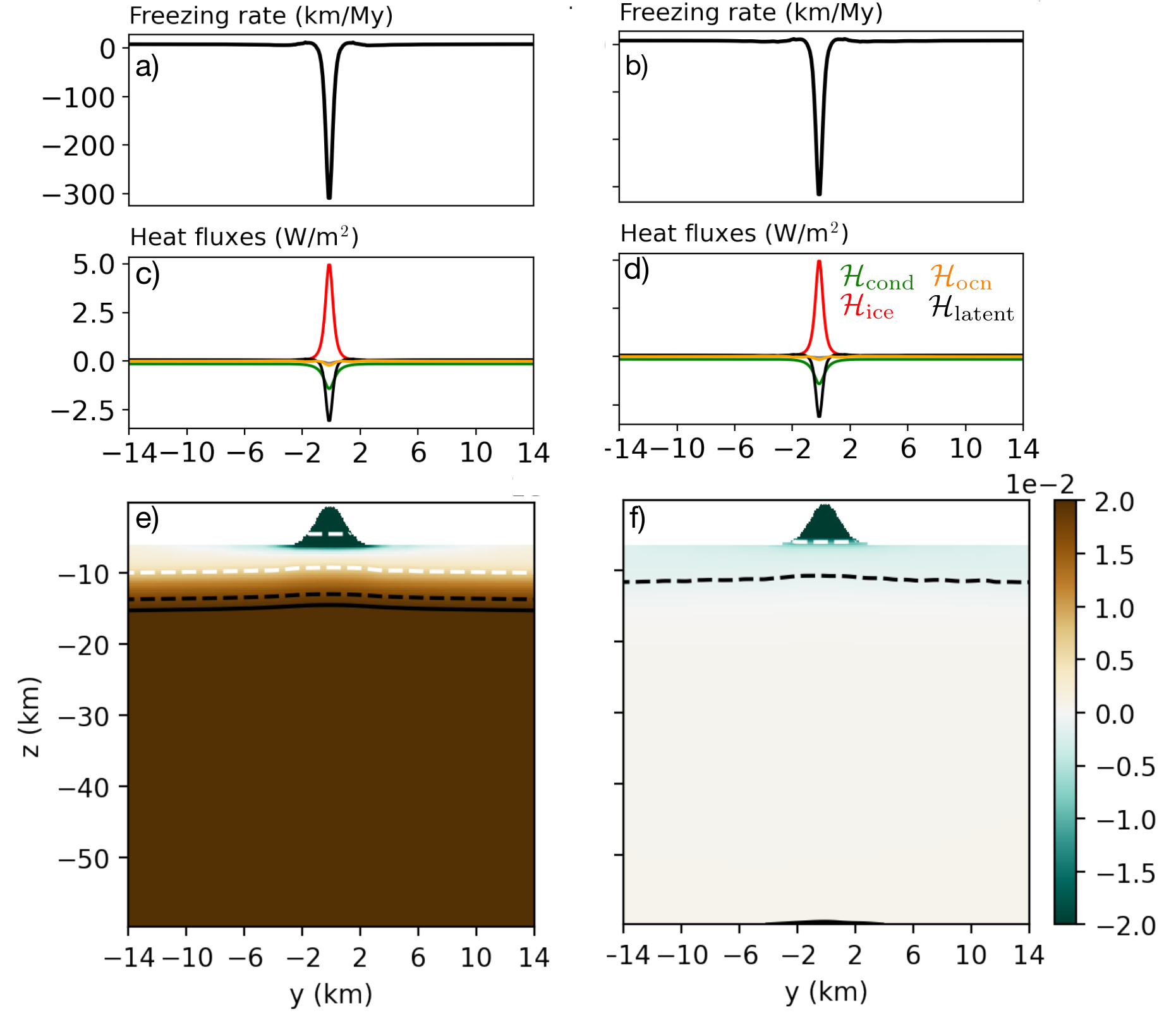}
    
    \caption{\small{Heat budget and ocean circulation in the shell-heating scenarios. Panel (a,b) show the freezing rate (left y-axis, black solid) and panel (c,d) show the heat budget terms (right y-axis, dashed curves) of the ice shell for S30c10 and S10c10, respectively. Panel (e,f) show salinity anomalies by shading and density anomalies by contours. Solid contour denotes positive density anomaly and negative contours denotes negative density anomaly. The black curves mark $\pm10^{-3}$~kg/m$^3$, and white curves mark $\pm10^{-2}$~kg/m$^3$. }}
    \label{fig:shell-heating}
  \end{figure*}

Given the difficulty in sustaining geysers in the core-heating scenarios, we now consider the case in which tidal heating is assumed to occur in the ice shell itself. The heating will be naturally amplified over the polar regions and therefore close to where geysers are observed (red curves in Fig.~\ref{fig:shell-heating}c,d). This is because thinner ice tends to enhance deformation and hence dissipation. Furthermore, with enhanced heating, the geyser regions becomes warmer and thus more prone to deformation \citep{Beuthe-2019:enceladuss}. This so-called rheology feedback, focuses tidal heating even more strongly than conductive heat loss (green curves), directly contributing to geyser sustenance. The state of affairs is presented in Fig.~\ref{fig:shell-heating}a,b).

\subsection{Plume dynamics and tracer transport timescales}
\label{sec:tracer-transport}
\begin{figure*}
    \centering
    \includegraphics[page=13,width=\textwidth]{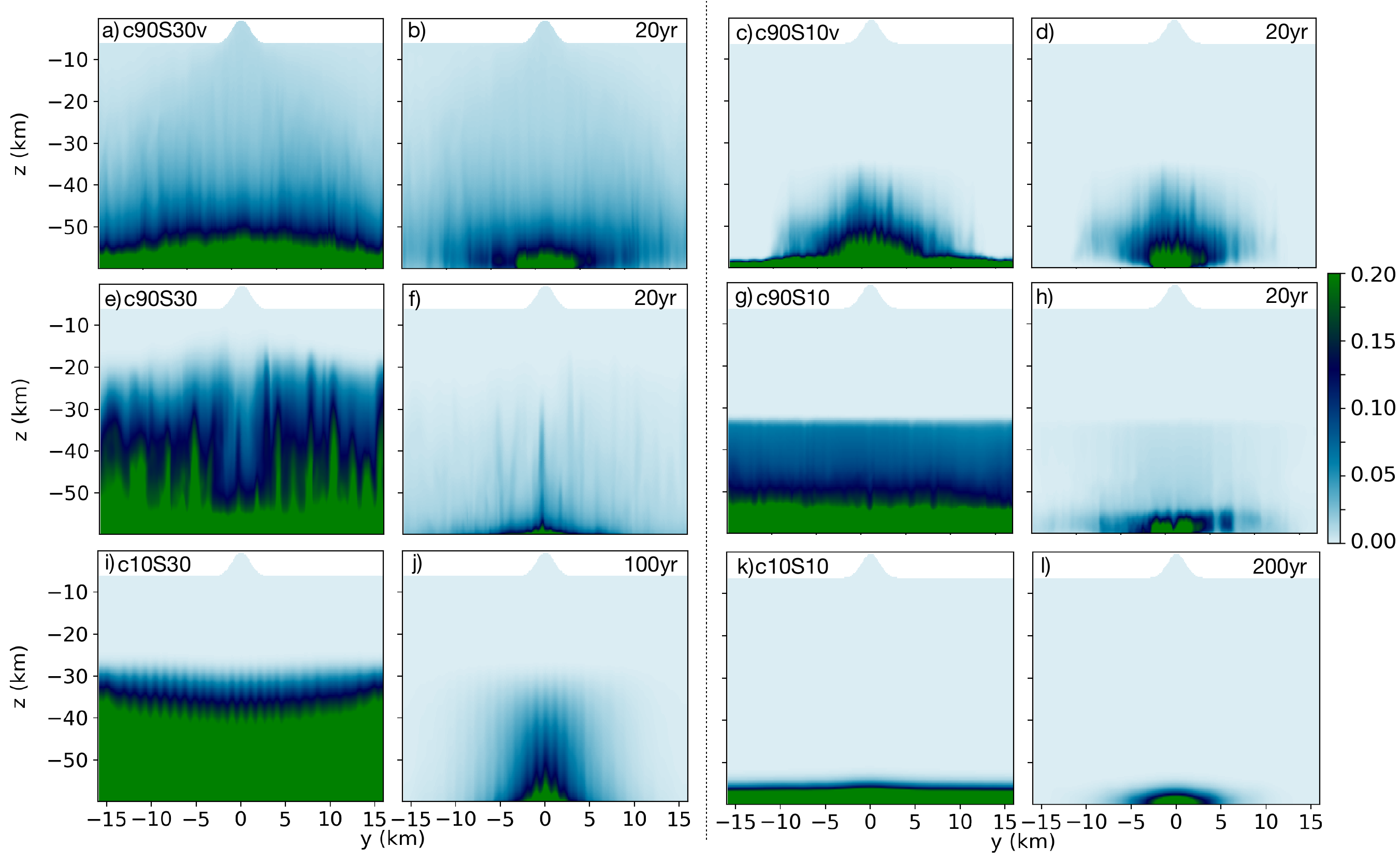}
    
    \caption{\small{Tracer distribution by the end of simulations. Tracers are released uniformly from the entire seafloor in panel (a,c,e,g,i,k), and are released only from a narrow zone on the seafloor right under the geyser in panel (b,d,f,h,j,l). The left two columns, from top to bottom are results for S30c90v, S30c90, S30c10. The right two columns, from top to bottom are shown for S10c90v, S10c90, S10c10. The lengths of integration since tracers are released are marked on the right for each scenario.}}
    
    \label{fig:tracer}
  \end{figure*}

Future sensors are likely to be deployed above the ice, and so it is important to come to some understanding of likely bottom-to-top transport time scales. This informs us of how long chemical components, introduced near the seafloor \citep{Parkinson-Liang-Yung-et-al-2008:habitability, Glein-Baross-Waite-2015:ph, McKay-Davila-Glein-et-al-2018:enceladus}, will have time to react with the ocean before reaching the underside of the ice and thence ejected out in a geyser. To approach this question, we release two types of passive tracer: one which is uniform in space at the seafloor and the other localised in space, at $y=0$, right under the geyser. The concentration of the tracer is restored to 1 at the bottom of the model, mimicking a chemical/biological process that keeps the tracer concentration at a certain level. After 20-200 years \footnote{We integrate longer for cases that have long transport timescale.}, the tracer distributions found in the six scenarios are shown in Fig.~\ref{fig:tracer}. As we now discuss, the tracer transport timescale is typically hundreds of years, if not longer, but can vary significantly depending on the ocean salinity and core-shell heat partition.

  Tracer transport is governed by ocean dynamics which responds to both the prescribed bottom heating and the salinity flux associated with freezing/melting of the ice above. Strong bottom heating (as in S30c90, S30c90v, S10c90, S10c90v) is expected to trigger stronger convection and hence more efficient transport compared to those with weak bottom heating (S30c10, S10c10). Furthermore, ocean salinity determines the sign and the magnitude of the thermal expansion coefficient. The same bottom heating will induce much weaker convection, if at all, in a fresh ocean due to the anomalous expansion of water near the freezing point. Thus very general considerations suggest that S30c90v and S30c90 should facilitate the most efficient tracer transport.

  \paragraph{Core-heating + salty ocean (S30c90v and S30c90).}
  
The tracer concentrations after 20 years of integration are presented in Fig.~\ref{fig:tracer}a,e for the two core-heating and salty-ocean scenarios. Even in these optimal scenarios, tracers have only been transported 10~km above the seafloor after 20 years. The tracer transport efficiency varies with the horizontal distance from the geyser (S30c90v and S30c90), indicating that the very short transport timescale proposed by \citet{Hsu-Postberg-Sekine-et-al-2015:ongoing} may be hard to achieve without extra buoyancy provided by, for example, gas bubbles.

In the localised heat source experiment, S30c90v, one would expect the tracer transport to only occur near $y=0$, where the heat source is located. However, as discussed above, baroclinic eddies grow along the edge of the heating line due to the temperature contrast between the hot plume and the cold ambient water \citep{Saunders-1973:instability, Visbeck-Marshall-Jones-1996:dynamics, Legg-Jones-Visbeck-1996:heton, Jacobs-Ivey-1999:rossby, Jacobs-Ivey-1998:influence}. These eddies induce turbulence and facilitate strong lateral mixing, leading to the almost uniform vertical tracer transport seen in Fig.~\ref{fig:tracer}a.

\begin{figure*}
    \centering
    \includegraphics[page=12,width=\textwidth]{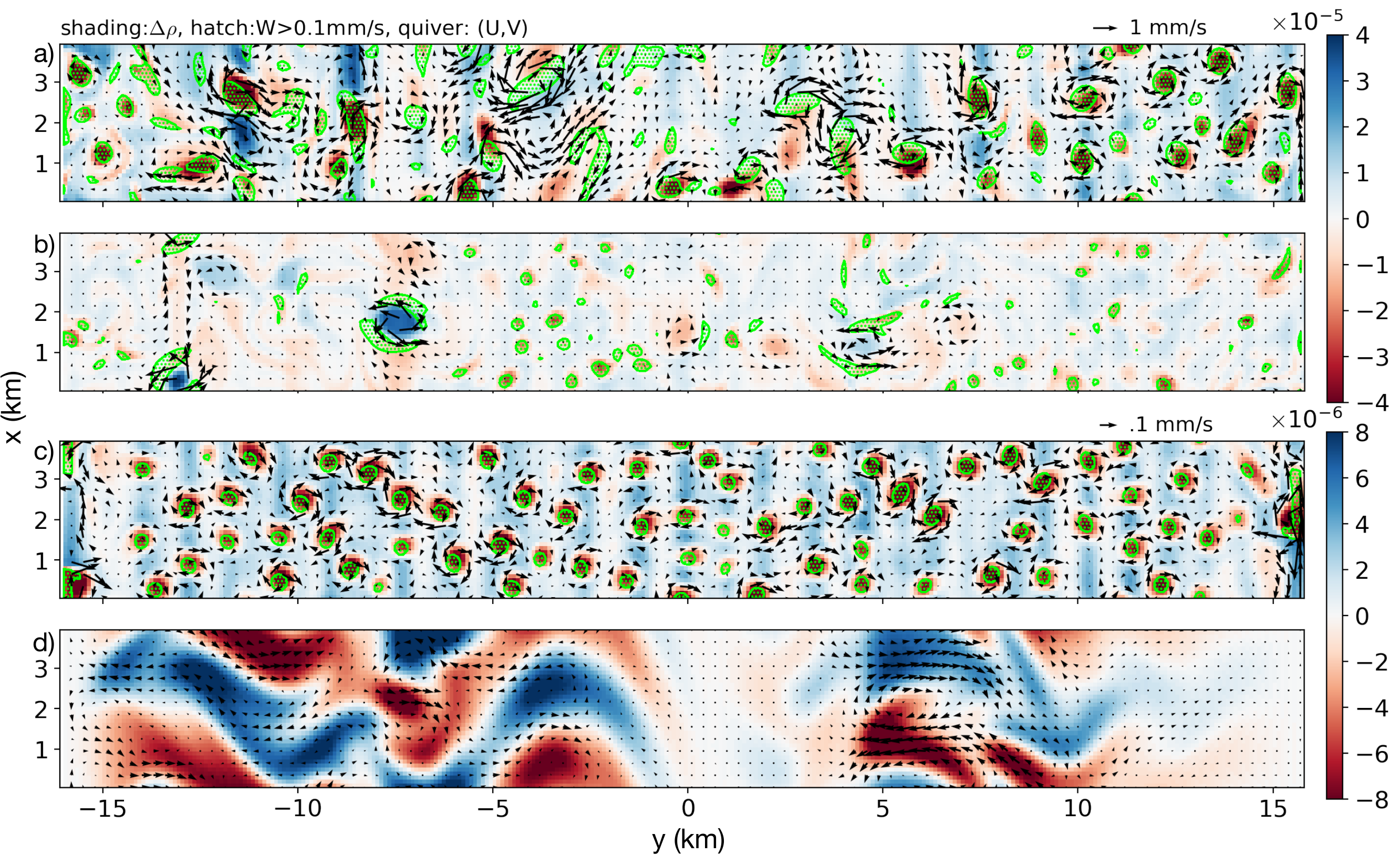}
    
    \caption{\small{Ocean dynamics close to the bottom for the uniform bottom heating scenarios and the shell heating scenarios. These plots are the same as the panel (e) of Fig.~\ref{fig:S30c90v-dynamics}. From top to bottom shows for S30c90, S10c90, S30c10 and S10c10. }}
    
    \label{fig:other-bottom-dynamics}
  \end{figure*}

In contrast, with the homogeneous bottom heating in S30c90, there is no externally forced temperature gradient to generate baroclinic eddies. Instead the convective plumes become organized into cones, surrounded by an azimuthal ``rim current''. This phenomenon has been well studied experimentally and numerically \citep{Jones-Marshall-1993:convection, Goodman-Collins-Marshall-et-al-2004:hydrothermal}. The scale over which the plumes congregate is given by the ``cone scaling''
  \begin{equation}
    \label{eq:cone-scaline}
    l_{\mathrm{cone}}=1.4 D^{1/2} \left( B_A /f^3\right)^{1/4},
  \end{equation}
where $D$ is the domain depth. The configuration of S30c90 yields a cone size of $l_{\mathrm{cone}}=300$~m, which is broadly consistent with the scales seen in Fig.~\ref{fig:other-bottom-dynamics}a. Fingerprints of the clustering of the convectively-modified fluid in to cones are visible in the tracer distribution seen in Fig.~\ref{fig:tracer}e).

Of the six scenarios considered here, only S30c90v and S30c90 are convectively unstable under the geyser due to the influence of salinity effects, as shown by the density contours in Fig.~\ref{fig:S30c90v-dynamics}b and Fig.S5b. The lower part of the ocean convects upward driven by the bottom heating whilst, at the same time, the upper ocean convects downward forced by the salinity flux associated with the freezing of the geyser (Fig.~\ref{fig:S30c90v-dynamics}c, Fig.S5f). In S30c90, the upward tracer transport by thermal convection seems to be inhibited in the vicinity of the geyser due to downward, salinity-driven convection quenching it from above. Instead, in S30c90v, tracer transport is almost uniform due to strong lateral mixing by turbulence. 

  \paragraph{Core-heating + fresh ocean (S10c90 and S10c90v).}
  
In a fresh ocean, the near bottom dynamics remain qualitatively similar to the high salinity case, but become less active (see Fig.~\ref{fig:S10c90v-dynamics}f and Fig.~S3i). This is because the same heat flux now induces a weaker buoyancy flux due to the much lower thermal expansion coefficient. The less vigorous dynamics is not only manifested by a weaker density gradient and a weaker ``rim current'', but also scales are reduced by a factor of 2-3. According to Eq.~(\ref{eq:deformation-radius}) and Eq.~(\ref{eq:cone-scaline}), this is broadly consistenct with a 30 fold reduction of $\alpha$ (the mean $\alpha$ of the convective layer is $3\times10^{-5}$/K in experiment S30c90 and $1\times10^{-6}$/K in S10c90). 

An interesting phenomenon observed in the fresh case is that, along with the upward-directed turbulent plumes, there are well-organized cold plumes shooting downwards, whose size is even larger than the upwelling convection (see Fig.~\ref{fig:S10c90v-dynamics}f). The same phenomenon is also seen in the uniform bottom heating case (Fig.~\ref{fig:other-bottom-dynamics}b). Irrespective of whether bottom heating is concentrated, in equilibrium the net upward heat flux should be the same at all levels. In other words, the upper surface must cool at the same rate as the bottom is warmed, and so upward and downward convective plumes should, in the net, be in balance with one-another. However, in our simulations, the downwelling plumes appear to be much larger and stronger. This asymmetry may be related to the strong temperature gradient in the upper part of the ocean and the depression of anomalous expansion with pressure. Near the upper boundary of the convective layer ($z=-34$~km), the perturbation associated with convective turbulence creates, for example, a cold bubble which sinks into the convective layer. The deeper it gets, the greater is the pressure leading to a more positive $\alpha$ and hence faster downward acceleration. As a result, this cold bubble will sink with increasing acceleration until it hits the bottom. In contrast, the temperature gradient near the seafloor is weak and convective plumes shooting upward will be weakened by the decrease of $\alpha$ and eventually suppressed once $\alpha$ becomes negative.

The ocean is stably stratified and diffusive in its upper regions. This stratified layer is a consequence of the anomalous expansion of fresh water. Forced by strong bottom heating, temperature increases approaching the seafloor (Fig.~\ref{fig:S10c90v-dynamics}a and Fig.S2a). However, such a temperature gradient stabilizes the upper part of the ocean, because fresh water contracts upon warming ($\alpha<0$) at low pressure near the freezing point. As pressure increases with depth, the temperature range over which $\alpha$ is negative gradually shrinks, and eventually vanishes around $z=-30$~km (the critical level). Only below this level does water expands upon warming allowing convective plumes to be triggered (Fig.~\ref{fig:S10c90v-dynamics}f and Fig.~\ref{fig:other-bottom-dynamics}c). Transporting tracers across the stratified layer may be a challenge due to the sluggish dynamics. This is reflected by the very low tracer concentrations observed above the critical level. If transport only occurs through vertical diffusion, and assuming a vertical diffusivity of $0.001$~m$^2$/s \citep{Kang-Mittal-Bire-et-al-2021:how}, 10~kyr is needed to diffuse through a 20~km layer \footnote{The diffusive timescale is $\tau_{\mathrm{diff}}=H^2/\kappa_v$, where $\kappa_v=0.001$~m$^2$/s is the vertical diffusivity and $H$ the vertical scale.}.

  \paragraph{Shell-heating.}

When the heating is primarily in the ice shell, the dynamics in the lower part of the ocean becomes even weaker. As shown in Fig.~\ref{fig:other-bottom-dynamics}(c), the combination of shell heating and a salty ocean (S30c10) facilitate the same columnar dynamics as S30c90, with buoyant fluid convecting upward under rotational control. Since the bottom heat flux here is $9$ times smaller than S30c90, the plume size is about half as large, as can be seen by comparing Fig.~\ref{fig:other-bottom-dynamics}c with Fig.~\ref{fig:other-bottom-dynamics}a. By year-20, tracers are only advected some 10~km above the seafloor (see Fig.S2e), and by year-100, they have only reached 30~km, half-way across the depth of the ocean (Fig.~\ref{fig:tracer}i).

If, by contrast, the ocean is fresh (S10c10), the entire ocean is stably stratified (Fig.~\ref{fig:shell-heating}f). Unlike in S90c10, there is no convective layer near the bottom (Fig.~\ref{fig:other-bottom-dynamics}d) because the top-to-bottom temperature gradient required to conduct the weak bottom heating away ($18.7$~mW/m$^2$) is only 0.2~K (assuming $\kappa_v=10^{-3}$~m$^2$/s), not enough to make the bottom $\alpha$ positive despite the high pressure there. In the absence of convection, vertical tracer transport is almost negligible. By the end of 200~years of integration, the tracers have only traveled 4~km (see Fig.~\ref{fig:tracer}kl), broadly consistent with a diffusion timescale.

In the upper part of the ocean, both of our shell-heating scenarios have a stably stratified layer, because the fresh water produced by ice melting near the geyser fills up the indentation beneath the geyser and diffuses downward, forming a freshwater lens. Again, this stratified layer may extend the tracer transport timescale to thousands of years, if vertical diffusion is the only process facilitating transport. Finally, we note that the stratified layer is not completely quiescent. Horizontal density gradients induced by the inhomogeneous melting trigger gravity waves, as can be seen in Figs.~S1h and S2h. 

To conclude, none of our six scenarios transports a significant amount of tracer to the geyser by the end of the simulations. Even with the help of convective plumes powered by strong bottom heating (S30c90, S30c90v), tracers are still 10-20~km away from the geyser by year 20. Tracer transport is even slower when bottom heating is weaker and/or ocean salinity is low. S30c10 transports tracers 10-20~km above the seafloor by year-100. In S10c90, since the upper part of the ocean is strongly stratified, the tracer transport is halted half-way and is unlikely to reach any farther. In S10c10, the tracers almost completely dwell at the seafloor because the ocean is stratified and quiet.

In our S30c90v and S10c90v experiments, we also released a second type of tracer which is only produced at $y=0$, right under the geyser where the bottom heat flux is concentrated. As shown in Fig.~\ref{fig:tracer}(a-d), near $y=0$ the concentration of this second tracer is comparable to that of the first, indicating that the tracer that reaches the geyser may be mostly from the vent area, and that the tracer transport timescale does not change significantly if the tracer is only produced near the vent. This initially concentrated tracer is carried out in all our experiments for reference.
 
  \section{Discussion}
  \label{sec:discussion}
We have investigated how ocean dynamics and tracer/heat transport near the south polar geysers of Enceladus can be affected by ocean salinity, the partition of heat between core and the shell and the heating distribution at the seafloor. Two key questions have motivated our study:

  \begin{enumerate}
  \item What are the conditions most likely to result in the geyser being self-sustained?

    We find that only when the ice shell produces most of the heat (S10c10, S30c10), can that heat be focused near the geyser, preventing it from freezing up. If the heat is primarily from the core, even if it is perfectly aligned with the geyser (as in S30c90v and S10c90v), convective plumes become baroclinically unstable, leaking most of the heat away from the geyser regions before reaching the ice. As a result, the geyser will freeze and close up.

\item How long does it take for tracers to be transported from the seafloor to the geyser region?

  Scenarios in which bottom-induced convection occurs over the entire ocean (S30c90, S30c90v) may be able to transport a significant amount of tracer to the geyser within a hundred years or so. Other scenarios either cannot transport tracer at all (S10c10) or the transport only reaches half-way (S10c90, S10c90v, S30c10) due to the stratification in the upper part of the ocean. Diffusion is the only process that can transport tracers across the stratified layer, but only on timescales of hundreds of years. 
  
  \end{enumerate}
  
  Given that a bottom-concentrated heat flux cannot remain concentrated as it is carried through the ocean, active polar geysers are likely driven by heating in the ice local to the geyser. Since the total heat loss through the geyser by far dominates other regions \citep{Howett-Spencer-Pearl-et-al-2011:high}, the ice shell has to be the major heat source. Therefore, amongst all the scenarios considered here, S10c10 and S30c10 are likely the most relevant. These suggest that tracer transport timescales are many hundreds of years. This is far greater than the month-to-year transport timescale suggested by \citet{Hsu-Postberg-Sekine-et-al-2015:ongoing}, and broadly aligns with the $\gg 100$~years estimate given by \citet{Zeng-Jansen-2021:ocean}. If this is the case then, when interpreting the chemical composition of ejecta, one has to account for the reaction of possible biosignatures emanating from the bottom with the ocean on their way up to the surface.

Many processes are absent from our study. First, we ignore the impact of erupting plumes upon the ocean. Based on the measurements of the Ultraviolet Imaging Spectrometer \citep{Hansen-Shemansky-Esposito-et-al-2011:composition}, the vapor production rate is estimated to be around 200~kg/s in total. Such eruptions have three major impacts: they drive flow toward the geyser to fill up the vacuum, a heat  source is required, and local salinity is increased because salt is left behind. However, the impact of eruptions is likely to be small. The flow speed required to compensate a mass sink of 200~kg/s over a total length of $\sim$420~km \footnote{Each geyser is around 130~km long and there are four of them \citep{Porco-Helfenstein-Thomas-et-al-2006:cassini}.} is only 0.04~mm/s, far weaker than the dynamics found here, even if the mass convergence is constrained to the top 10~m of the ocean. The associated salinity flux is only equivalent to freezing at a rate of 4~km/My over a 3-km wide geyser. This is again far smaller than the freezing/melting rate found in our simulations.
  
Second, we ignore the thermodynamical and dynamical interaction between the south pole and the rest of the ocean and assume that the heat budget is local. In reality, the south polar ocean is fully connected with the broader ocean. Forced by inhomogeneous heat and salinity fluxes from the ice, a meridional circulation could form that transports heat equatorward \citep{Kang-Mittal-Bire-et-al-2021:how}. This equatorward heat convergence cannot exceed the conductive heat loss rate through the equatorial ice shell. Observation tell us that the heat flux passing through the geysers completely dominates the heat flux anywhere \citep{Porco-Helfenstein-Thomas-et-al-2006:cassini, Howett-Spencer-Pearl-et-al-2011:high, Spencer-Howett-Verbiscer-et-al-2013:enceladus, Iess-Stevenson-Parisi-et-al-2014:gravity}. This suggests that the heat budget over the south pole of Enceladus may be roughly in balance. That said, future studies should explore the role of localised geysers in the general circulation.

\section*{Acknowledgements}
This work is carried out in the Department of Earth, Atmospheric and Planetary Science (EAPS) in MIT. WK acknowledges support as a Lorenz-Houghton Fellow supported by endowed funds in EAPS. JM acknowledges part-support from NASA Astrobiology Grant 80NSSC19K1427 “Exploring Ocean Worlds”. We all thank “Exploring Ocean Worlds” for helpful supports and discussions.

\section*{Data Availability}

 The supporting information provides detailed description of our model setup. We would like to provide more code and data upon reasonable request.



\bibliographystyle{mnras}
\bibliography{export} 




\appendix



\section{Description of the Ocean Model}
\label{sec:model-description}
To study small-scale ocean dynamics and transport timescales near the south polar geysers on Enceladus, we carry out a set of high-resolution non-hydrostatic simulations using the state-of-the-art Massachusetts Institute of Technology OGCM (MITgcm)\citep{MITgcm-group-2010:mitgcm, Marshall-Adcroft-Hill-et-al-1997:finite}. The domain is a periodic channel bounded by walls to the north and south, of length 4~km in $x$ and width 32~km in $y$. The geyser stripe lies along $x$ in the middle of the channel. This is broadly motivated by the observation that the ``tiger stripes'' are some four hundred kms long and spaced 30-35~km apart \citep{Porco-Helfenstein-Thomas-et-al-2006:cassini}. To reduce computational cost we simulate a narrow range in $x$ given that the dynamics are likely to be homogeneous along the stripe. A horizontal resolution of 100-meter is used to capture the small-scale convective and turbulent motions.

The vertical extent of the domain is 60~km and comprises the ocean and ice shell of Enceladus \citep{Iess-Stevenson-Parisi-et-al-2014:gravity, McKinnon-2015:effect}. It is divided into 200 evenly spaced layers, each 300~m thick. The ice thickness $H$ is set to $H_0=7$~km except for the geyser regions, where a Gaussian-shaped indentation --- an upside down trench --- pushes upward into the ice shell from below. At the center of the trench, the ice thickness diminishes to 500~m, and the trench is around 2~km wide, broadly consistent with the observations \citep{Hemingway-Mittal-2019:enceladuss, Porco-Helfenstein-Thomas-et-al-2006:cassini}.
\begin{equation}
  \label{eq:Hice}
  H=H_0-\Delta H \exp\left(-\frac{y^2}{2\sigma_H^2}\right), ~~~~ (-16 \leq y\leq 16~\mathrm{km})
\end{equation}
where $\Delta H=6.5$~km and $\sigma_H=1$~km. 

Since the domain is near the south pole and is rather small compare to the size of Enceladus ($a=252$km), the f-plane assumption is made. The Coriolis parameter is set to $f_0=2\omega=1.07\times10^{-4}$/s.  We account for the variation of gravity with depth
  \begin{equation}
    \label{eq:g-z}
    g(z)=\frac{G\left[M-(4\pi/3)\rho_{\mathrm{out}}(a^3-(a-z)^3)\right]}{(a-z)^2},
  \end{equation}
where the outer layer density $\rho_{\mathrm{out}}$ is set to 1000~kg/m$^3$ and the Enceladus mass and radius are $M=1.08\times 10^{20}$~kg and $a=252$~km. $G=6.67\times10^{-11}$~N/m$^2$/kg$^2$ is the gravitational constant. We choose to use the ``MDJWF'' equation of state (EOS) \citep{McDougall-Jackett-Wright-et-al-2003:accurate}, which has been shown to be a good nonlinear fit of the full TEOS-10 EOS based on the Gibbs function analysis \citep{McDougall-Barker-2011:getting}.

Since tidal forcing and libration motions are not simulated in our model, we use explicit diffusivity to account for the induced mixing of heat and salinity. According to \citep{Rekier-Trinh-Triana-et-al-2019:internal}, the tidal dissipation in the ocean is mostly induced by the libration motion and the global dissipation rate $E$ should be around 1~MW. With this, and guided by terrestrial oceanography, we follow the review of \citep{Wunsch-Ferrari-2004:vertical} to estimate the vertical diffusivity 
  \begin{equation}
    \label{eq:kappav}
    \kappa_{v}=\frac{\Gamma \varepsilon}{\rho_0 N^{2}},
  \end{equation}
where $\Gamma\sim 0.2$ is the assumed efficiency that kinetic energy dissipation results on potential energy production. Here $\varepsilon=E/V$ is the ocean dissipation rate per area, $V=4\pi a^2D$ is the total volume of the Enceladean ocean, $\rho_0\sim 1000$~kg/m$^3$ is the water density and $N^2=g(\partial \ln \rho/\partial z)\sim g (\Delta \rho/rho_0)/D$ is the buoyancy frequency of the ocean, where $g$ and $D$ is gravity and ocean depth, respectively. $\Delta \rho/\rho_0$ can be estimated by $\alpha_T \Delta T_f$, where $\alpha_T\sim 1\times10^{-5}$/K is a typical thermal expansion coefficient, and $\Delta T_f\sim 0.1$K is the equator-to-pole temperature contrast \citep{Kang-Mittal-Bire-et-al-2021:how}. Substituting these values into Eq.\ref{eq:kappav}, we find that $\kappa_v\sim 0.001$~m$^2$/s. This is the diffusivity used in both the vertical and horizontal for temperature, salinity and passive tracers. The viscosity is set to $0.05$~m$^2$/s to remove grid-scale noise and maintain numerically stability. 
 
\subsection{Ice heat budget}
\label{sec:ice-heat-budget}
  The ice shell loses heat through conduction ($\mathcal{H}_{\mathrm{cond}}$), and gains heat by tidal dissipation ($\mathcal{H}_{\mathrm{ice}}$), gains/loses heat from the ice ($\mathcal{H}_{\mathrm{latent}}$) and from the ocean ($\mathcal{H}_{\mathrm{ocn}}$), the latter of which partly comes from the tidal heat produced in the core ($\mathcal{H}_{\mathrm{core}}$) and partly from ocean heat transport ($\mathcal{H}_{\mathrm{oht}}$). Since our focus here is the contrast between the geyser regions and the space in between, we assume that, averaged over the domain, the ice is neither freezing nor melting, $\overline{q}=0$, where $q$ is the freezing rate. This necessarily implies that the domain-averaged $\mathcal{H}_{\mathrm{latent}}$ vanishes. Furthermore, for simplicity we ignore heat transport from other latitudes $\mathcal{H}_{\mathrm{oht}}$. According to \citet{Kang-Mittal-Bire-et-al-2021:how}, $\mathcal{H}_{\mathrm{oht}}$ should be negative if heat is transported down-gradient from poles to equator. Thus, to ensure $\overline{q}=0$, local heat production in the ice and core must be greater than $\mathcal{H}_{\mathrm{cond}}$. However, the magnitude of $\mathcal{H}_{\mathrm{oht}}$ is likely to be small, because little heat can be lost through the thick equatorial ice shell \citep{Cadek-Soucek-Behounkova-et-al-2019:long, Kang-Mittal-Bire-et-al-2021:how} placing a limit on $\mathcal{H}_{\mathrm{oht}}$. We now present the formula for $\mathcal{H}_{\mathrm{cond}}$, $\mathcal{H}_{\mathrm{ice}}$ and $\mathcal{H}_{\mathrm{core}}$.

The heat conduction through the ice shell, $\mathcal{H}_{\mathrm{cond}}$, is induced by the temperature gradient between the surface $T_s$ and the water-ice interface $T_f$ (freezing point). To compute $\mathcal{H}_{\mathrm{cond}}$, we first solve a 1D steady-state heat conduction model,
\begin{eqnarray}
  \label{eq:heat-condution}
  \frac{\partial}{\partial z}\left(\kappa \frac{\partial T}{\partial z}\right)=0.
\end{eqnarray}
  with fixed temperature at the top and bottom boundaries to obtain a vertical temperature profile. In the above equation, $\kappa$, the heat conductivity of ice, varies inversely with the temperature \citep{Slack-1980:thermal,Petrenko-Whitworth-1999:physics}
\begin{equation}
  \label{eq:kappa-T}
  \kappa(T)=\kappa_0/T.
\end{equation}
Given a temperature profile, the conductive heat flux can be estimated from
\begin{equation}
  \mathcal{H}_{\mathrm{cond}}=\frac{\kappa_{0}}{H} \ln \left(\frac{T_{f}}{T_{s}}\right),
  \label{eq:H-cond}
\end{equation}
if the thickness of the ice $H$ (Eq.\ref{eq:Hice}), surface temperature $T_s=59$K and freezing point $T_f\sim 273$K are known. Substituting typical numbers we find that the domain-averaged $\mathcal{H}_{\mathrm{cond}}$ is around 185~mW/m$^2$.

The ice dissipation $\mathcal{H}_{\mathrm{ice}}$ should be constant if there are no ice thickness variations, because the domain is tiny compared to Enceladus' size and the ice dissipation rate only varies on the large scales \citep{Beuthe-2018:enceladuss}. However, in the presence of thickness gradients (Eq.\ref{eq:Hice}), regions with thinner ice will be more mobile and thereby produce more heat. Following \citet{Beuthe-2019:enceladuss}, we account for this rheology feedback by multiplying through by a thickness-dependent factor
\begin{equation}
  \label{eq:H-tide}
  \mathcal{H}_{\mathrm{ice}}=\overline{\mathcal{H}_{\mathrm{ice}}}\frac{(H/\overline{H})^{p_\alpha}}{\overline{(H/\overline{H})^{p_\alpha}}},
\end{equation}
where $\overline{(\cdot)}$ denotes the domain average and the ice thickness $H$ is given by Eq.\ref{eq:Hice}.

Heat production in the core, $\mathcal{H}_{\mathrm{core}}$, is set to a constant throughout the domain when distriubuted heating is used. Some of our calculations assume that bottom heat flux is concentrated in a narrow stripe which is perfectly aligned with the geyser topography thus:
\begin{equation}
  \label{eq:H-core-vent}
  \mathcal{H}_{\mathrm{core}}^{\mathrm{vent}}=\frac{\overline{\mathcal{H}_{\mathrm{core}}}}{\sqrt{2\pi} \sigma_{\mathrm{core}}}\exp\left(-\frac{y^2}{2\sigma_{\mathrm{core}}^2}\right),
\end{equation}
where $\overline{\mathcal{H}_{\mathrm{core}}}$ is the domain-averaged bottom heat flux which is set to the core-heating percentage $c$ times the domain-averaged conductive heat loss rate $\overline{\mathcal{H}_{\mathrm{cond}}}$ (see Eq.\ref{eq:H-cond}) and $\sigma_{\mathrm{core}}$ is the standard deviation of the assumed Gaussian profile.
Setting $\sigma_{\mathrm{core}}=\sigma_{\mathrm{H}}/5=200$~m, yields a maximum heat flux of around $10$~W/m$^2$, broadly consistent with the calculations of \citet{Choblet-Tobie-Sotin-et-al-2017:powering}.

  \subsection{Water-ice interaction}
\label{sec:water-ice-interaction}
  The interaction between the ocean and ice is simulated using MITgcm's ``shelf-ice'' package \citep{Losch-2008:modeling, Holland-Jenkins-1999:modeling}. We turn on the ``boundary layer'' option to enhance the behavior near steep ice topographies. The code is modified to account for a gravitational acceleration different from that on earth, the temperature dependence of heat conductivity, and the extra heating due to ice dissipation. The ice is allowed to freeze/melt in response to the heat deficit/gain of the ice layer just above the ocean. The heat budget includes 1) the heat transmitted upward by ocean $\mathcal{H}_{\mathrm{ocn}}$, 2) the heat loss through the ice shell due to heat conduction $\mathcal{H}_{\mathrm{cond}}$ (Eq.\ref{eq:H-cond}), 3) the tidal heating generated inside the ice shell $\mathcal{H}_{\mathrm{ice}}$ (Eq.\ref{eq:H-tide}) and 4) the latent heat release $\mathcal{H}_{\mathrm{latent}}$. Following \citet{Holland-Jenkins-1999:modeling} and \citet{Losch-2008:modeling}, we write the equivalent ``three equation formula'' for our system. First, the heat and salinity budgets suggest
  \begin{eqnarray}
    &~&\mathcal{H}_{\mathrm{ocn}}-\mathcal{H}_{\mathrm{cond}}+\mathcal{H}_{\mathrm{tidal}}=-L_fq-C_p(T_{\mathrm{ocn-top}}-T_b)q\label{eq:boundary-heat}\\
&~&\mathcal{F}_{\mathrm{ocn}}=-S_bq-(S_{\mathrm{ocn-top}}-S_b)q, \label{eq:boundary-salinity}   
  \end{eqnarray}
  where $T_{\mathrm{ocn-top}}$ and $S_{\mathrm{ocn-top}}$ denote the temperature and salinity in the top grid of the ocean\footnote{When model resolution is smaller than the boundary layer thickness, the salinity below the upper-most grid cell also contributes to $T_{\mathrm{ocn-top}}$ and $S_{\mathrm{ocn-top}}$.}, $T_b,\ S_b$ denote the temperature and salinity in the ``boundary layer'', and $q$ denotes the freezing rate in $kg/s$. $C_p=4000$~J/kg/K is the heat capacity of the ocean, $L_f=334000$~J/kg is the latent heat of fusion of ice. By assumption, the boundary layer temperature equals the local freezing point,
\begin{equation}
    \label{eq:freezing-point}
    T_b=T_f(S_b,P)=c_0+b_0P+a_0S_b,
  \end{equation}
  where constants $a_0=-0.0575$~K/psu, $b_0=-7.61\times10^{-4}$~K/dbar and $c_0=0.0901$~degC. The pressure $P$ can be calculated using hydrostatic balance $P=\rho_igH$ ($\rho_i=917$~kg/m$^3$ is the density of the ice and $H$ is the ice thickness).

  Then, the ocean-ice heat exchange $\mathcal{H}_{\mathrm{ocn}}$ and salinity exchange $\mathcal{F}_{\mathrm{ocn}}$ in Eq.\ref{eq:boundary-heat} can be written as
  \begin{eqnarray}
    \mathcal{H}_{\mathrm{ocn}}&=&C_p(\rho_{0}\gamma_T-q)(T_{\mathrm{ocn-top}}-T_b),\label{eq:H-ocn}\\
    \mathcal{F}_{\mathrm{ocn}}&=&(\rho_{0}\gamma_S-q)(S_{\mathrm{ocn-top}}-S_b) \label{eq:S-ocn}
  \end{eqnarray}
  where $\gamma_T=\gamma_S=10^{-5}$~m/s are the exchange coefficients for temperature and salinity, and $T_b$ denotes the and temperature in the ``boundary layer''. The terms associated with $q$ are the heat/salinity change induced by the deviation of $T_{\mathrm{ocn-top}},\ S_{\mathrm{ocn-top}}$ from that in the ``boundary layer'', where melting and freezing occur. $T_b$ equals the freezing temperature $T_f(S_b,P)$ (Eq.\ref{eq:freezing-point}) at pressure $P$ and salinity $S_b$ by definition. $\rho_0$ is the reference density.

  Combining Eq.~(\ref{eq:boundary-heat}), Eq.~(\ref{eq:freezing-point}) and Eq.~(\ref{eq:boundary-salinity}), the two unknowns, $S_b$ and $q$, can be solved. When freezing occurs ($q>0$), salinity flux $\rho_{w0}\gamma_S(S_{\mathrm{ocn-top}}-S_b)$ is negative (downward). This leads to a positive tendency of salinity at the top of the model ocean, and meanwhile temperature will be relaxed toward the freezing point of the boundary layer.
  \begin{eqnarray}
    \frac{dS_{\mathrm{ocn-top}}}{dt}&=&\frac{-\mathcal{F}_{\mathrm{ocn}}}{\rho_{w0}\delta z}=\frac{1}{\rho_{w0}\delta z}(\rho_{w0}\gamma_S-q)(S_b-S_{\mathrm{ocn-top}})=\frac{qS_{\mathrm{ocn-top}}}{\rho_{w0}\delta z},\label{eq:S-tendency}\\
    \frac{dT_{\mathrm{ocn-top}}}{dt}&=&\frac{-\mathcal{H}_{\mathrm{ocn}}}{C_p\rho_{w0}\delta z}=\frac{1}{\rho_{w0}\delta z}(\rho_{w0}\gamma_T-q)(T_b-T_{\mathrm{ocn-top}})\nonumber\\
    &=&\frac{1}{C_p\rho_{w0}\delta z}\left[\mathcal{H}_{\mathrm{tidal}}-\mathcal{H}_{\mathrm{cond}}+L_fq+C_p(T_{\mathrm{ocn-top}}-T_b)q\right]
  \end{eqnarray}
  where $\delta z$ is the thickness of the ``boundary layer'' at the ocean-ice interface.

The flow speed is relaxed to zero at the top boundary at a rate of $1\times10^{-3}$~m/s.

\subsection{Model spin up}
\label{sec:spin-up}
Due to the high resolution of our model it is computationally expensive. Even when running efficiently on 256 CPUs, the S30c90v scenario can only achieve 3 model months per day. Given such clocktimes, integrating the water-ice system out to equilibrium over many scenarios is not possible. To accelerate convergence, we therefore first spin up our simulations in a 2D setup, in which all else is the same as the 3D model. After the temperature profile and freezing rate approaches equilibrium --- around 200-800 model-years, depending on the scenario --- we initialise the 3D model after adding small random perturbations to trigger instability. The 3D model is then integrated on for a further 10 years to develop a turbulent solution. Passive tracers are ``released'' from the seafloor. A few select integrations were also run out to equilibrium using only the 3D model without use of the 2D intermediate model.

The initial temperature and salinity profiles used to start the 2D integrations are prescribed analytically thus:
\begin{eqnarray}  
  T_i(\phi,z)&=&\min\left\{T_f(S_0,\rho_ig_0H_0)+b_0\rho_wg_0\max\{z+H_0,0\}+\Gamma_T \frac{\max\{-z-H_0,0\}}{D_0},~T_{c}\right\} \label{eq:Tini}\\
  S_i(\phi,z)&=&S_0+\Gamma_S \frac{\max\{z+H_0,0\}}{\Delta H}, \label{eq:Sini}
\end{eqnarray}
where $\rho_i,\ \rho_w$ are the densities of ice and water, $g_0$ is the surface gravity, $S_0$ is the assumed salinity, $H_0=7$~km and $D_0=53~$km are the ice thickness and ocean depth outside the geyser regions, and $\Delta H=6.5$~km is the depth of the trench. The freezing point $T_f$ and the pressure depression coefficient $b_0$ defined in Eq.\ref{eq:freezing-point}. $z$ denotes the height counting from the ice surface (positive upward). The above formula guarantees that the temperature at the top of the ocean is equal to the local freezing temperature $T_f(S_0,rho_ig_0H(\phi))$, where $H(\phi)$ is the ice thickness profile defined in Eq.\ref{eq:Hice}.


\begin{table*}[hptb!]
  
  \centering
  \begin{tabular}{lll}
    \hline
    Symbol & Name & Definition/Value\\
    \hline
    \multicolumn{3}{c}{Enceladus parameters}\\
    \hline
    $a$ & radius & 252~km\\
    $\delta$ & obliquity & 27$^\circ$\\
    $H_0$ & ice thickness for non-geyser regions & 7~km: ref \citep{Hemingway-Mittal-2019:enceladuss}  \\
    $D_0$ & ocean depth for non-geyser regions& 53~km: ref \citep{Hemingway-Mittal-2019:enceladuss} \\
    $\Delta H$ & depth of the geyser trench & 6.5~km\\
    $\sigma_H$ & width of the geyser trench & 1~km\\
        $\sigma_{\mathrm{core}}$ & width of vent in S30c90v and S10c90v & 200~m\\
    $\Omega$ & rotation rate & 5.307$\times$10$^{-5}$~s$^{-1}$\\
    $g_0$ & surface gravity & 0.113~m/s$^2$\\
    $T_s$ & ice surface temperature& 59K\\
    \hline
    \multicolumn{3}{c}{Physical constants}\\
    \hline
    $L_f$ & fusion energy of ice & 334000~J/kg\\
    $C_p$ & heat capacity of water & 4000~J/kg/K\\
    $T_f(S,P)$ & freezing point & Eq.\ref{eq:freezing-point}\\
    $\rho_i$ & density of ice & 917~kg/m$^3$ \\
    $\kappa_0$ & conductivity coeff. of ice & 651~W/m:  ref \citep{Petrenko-Whitworth-1999:physics}\\
    $p_\alpha$ & ice dissipation amplification factor & -2 $\sim$ -1 \\
    $\eta_m$ & ice viscosity at freezing point & 10$^{14}$~Ps$\cdot$s\\
    $T_f$ & freezing point & Eq.\ref{eq:freezing-point}\\
    \hline
    \multicolumn{3}{c}{Default parameters in the ocean model}\\
    \hline
    $\nu_h,\ \nu_v$ & horizontal/vertical viscosity & 0.05~m$^2$/s\\
    $\kappa_h,\ \kappa_v$ & horizontal/vertical diffusivity & 0.001~m$^2$/s\\
    $(\gamma_T,\ \gamma_S,\ \gamma_M)$ & water-ice exchange coeff. for T, S \& momentum & (10$^{-5}$, 10$^{-5}$, 10$^{-3}$)~m/s\\
    $g$ & gravity in the ocean & Eq.\ref{eq:g-z}\\
    $c$ & percentage of heat produced in the core & 0.1 or 0.9\\
    $S_0$ & ocean mean salinity & 10 or 30~psu (g/kg)\\
    $\mathcal{H}_{\mathrm{cond}}$ & conductive heat loss through ice & Eq.\ref{eq:H-cond}\\
    $\overline{\mathcal{H}_{\mathrm{ice}}}$ & domain-averaged ice dissipation & $(1-c)\overline{\mathcal{H}_{\mathrm{cond}}}$ \\
    $\overline{\mathcal{H}_{\mathrm{core}}}$ & domain-averaged bottom heat flux & $c\overline{\mathcal{H}_{\mathrm{cond}}}$  \\
    $\mathcal{H}_{\mathrm{ice}}$ & ice dissipation profile & Eq.\ref{eq:H-tide}  \\
    $\mathcal{H}_{\mathrm{core}}^{\mathrm{vent}}$& bottom heat flux used in S30c90v and S10c90v & Eq.\ref{eq:H-core-vent}\\
    \hline
     \end{tabular}
  \caption{Default model parameters. }
  \label{tab:parameters}
  
\end{table*}

\begin{figure*}
    \centering
    \includegraphics[page=1,width=\textwidth]{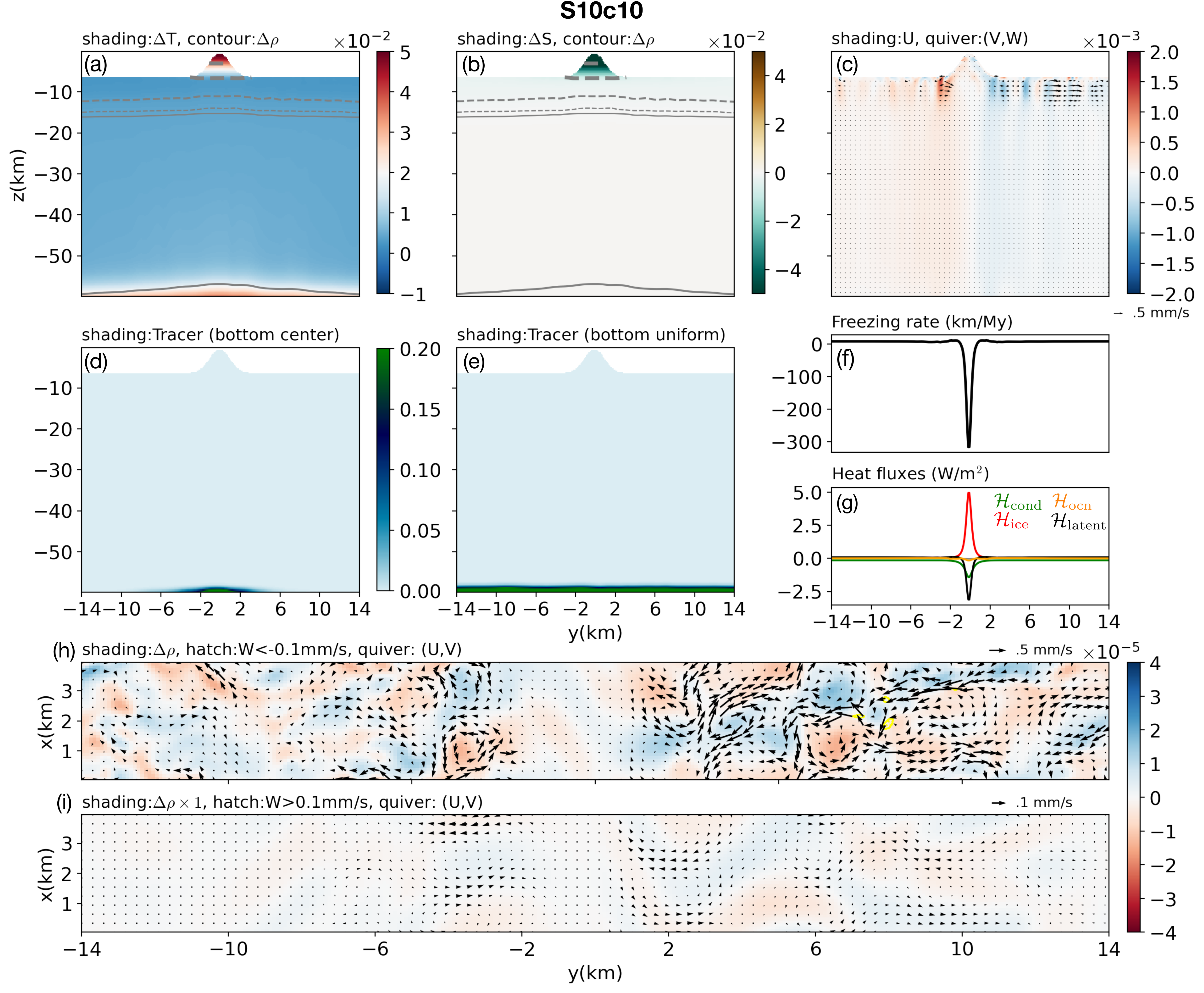}
    
    \caption{\small{Snapshot taken at the end of the simulation for the S10c10 scenario. Panel (a) shows the temperature anomalies at a given $x$ in shading and density anomalies in contours. Solid contour denote positive density anomaly and dashed contours denote negative density anomaly. The gray contours mark $\Delta \rho=\pm 10^{-4}$, $\pm 8\times 10^{-4}$~kg/m$^3$, $\pm 5\times 10^{-3}$~kg/m$^3$, $\pm 5\times 10^{-2}$~kg/m$^3$ as the line gets thicker. Panel (b) is similar to panel (a) except salinity is shown in place of temperature. Panel (c) shows the dynamics for a vertical cross-section, flow into and out of the presented plane in shading and flow speeds aligned with the plane in quivers. Panel (d,e) show the concentration of two tracers averaged in $x$, one released from $y=0$ at the bottom and the other released from the whole seafloor. Panel (f) shows the heat budget (dashed curves, right y-axis) and the freezing/melting rate (solid black curve, left y-axis) of the ice shell. Red, orange, green and black dashed curves, respectively, represent the ice dissipation $\mathcal{H}_{\mathrm{ice}}$, the heat absorbed from the ocean $\mathcal{H}_{\mathrm{ocn}}$, the conductive heat loss through the ice $\mathcal{H}_{\mathrm{cond}}$ and the latent heat release $\mathcal{H}_{\mathrm{latent}}$. The gray dashed curve shows the residue of the heat budget, i.e., $\mathcal{H}_{\mathrm{latent}}+\mathcal{H}_{\mathrm{ice}}+\mathcal{H}_{\mathrm{ocn}}-\mathcal{H}_{\mathrm{cond}}$, which is close to zero. Panel (g,h) show the dynamics in a horizontal plane, horizontal flow speeds in quivers, density anomaly in the shading, and areas with vertical speed beyond a certain threshold (see text just above the figure) are marked by hashes. The plane shown by panel (g) is taken just below the ice shell (z=-9km), and the plane shown by panel (h) is just above the seafloor (z=-54km).}}
    
    \label{fig:S10c10}
  \end{figure*}

  \begin{figure*}
    \centering
    \includegraphics[page=2,width=\textwidth]{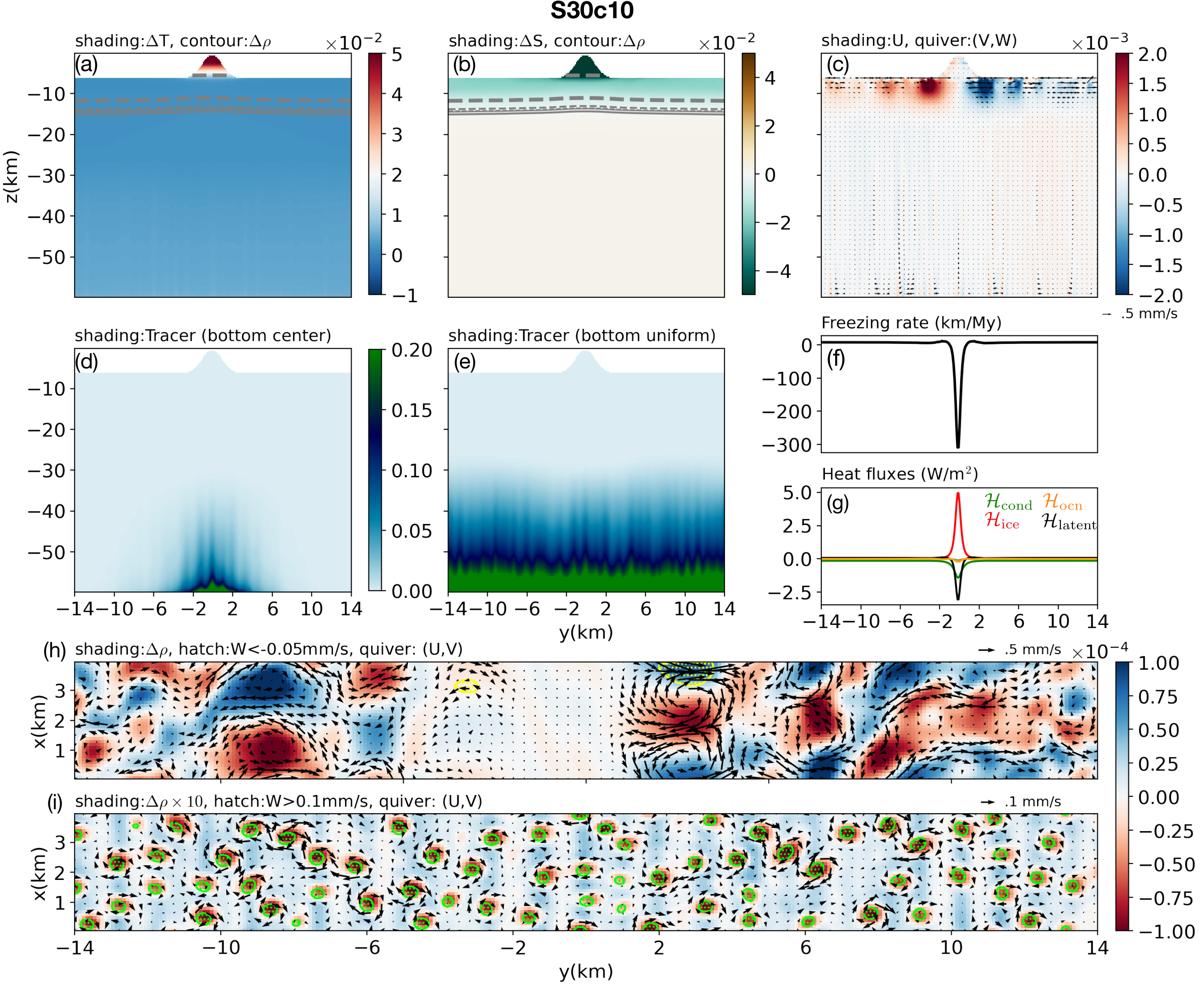}
    
    \caption{\small{Similar to Fig.~\ref{fig:S10c10}, except for the S30c10 scenario.}}
    
    \label{fig:S30c10}
  \end{figure*}

  \begin{figure*}
    \centering
    \includegraphics[page=3,width=\textwidth]{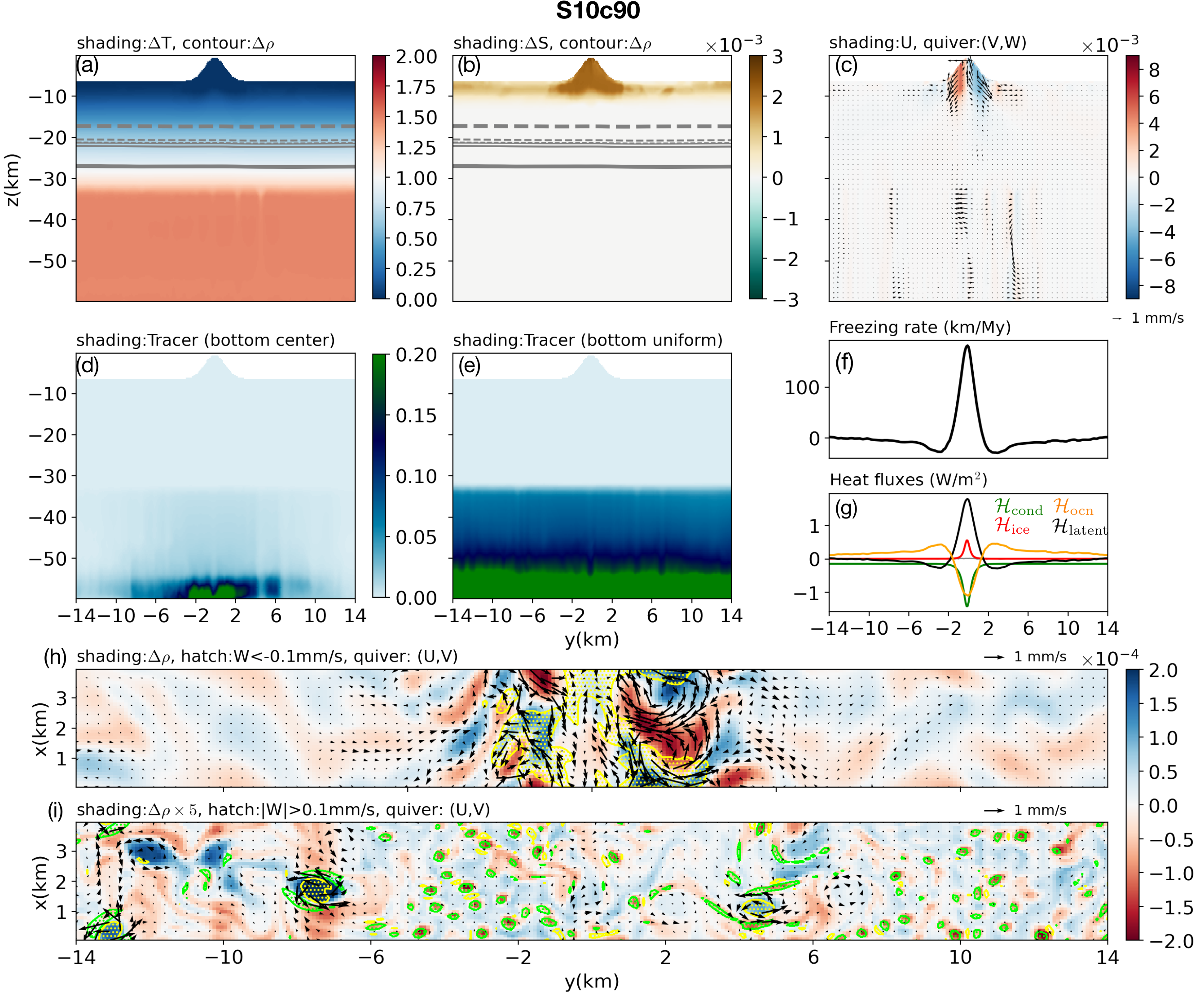}
    
    \caption{\small{Similar to Fig.~\ref{fig:S10c10}, except for the S10c90 scenario.}}
    
    \label{fig:S10c90}
  \end{figure*}

  \begin{figure*}
    \centering
    \includegraphics[page=4,width=\textwidth]{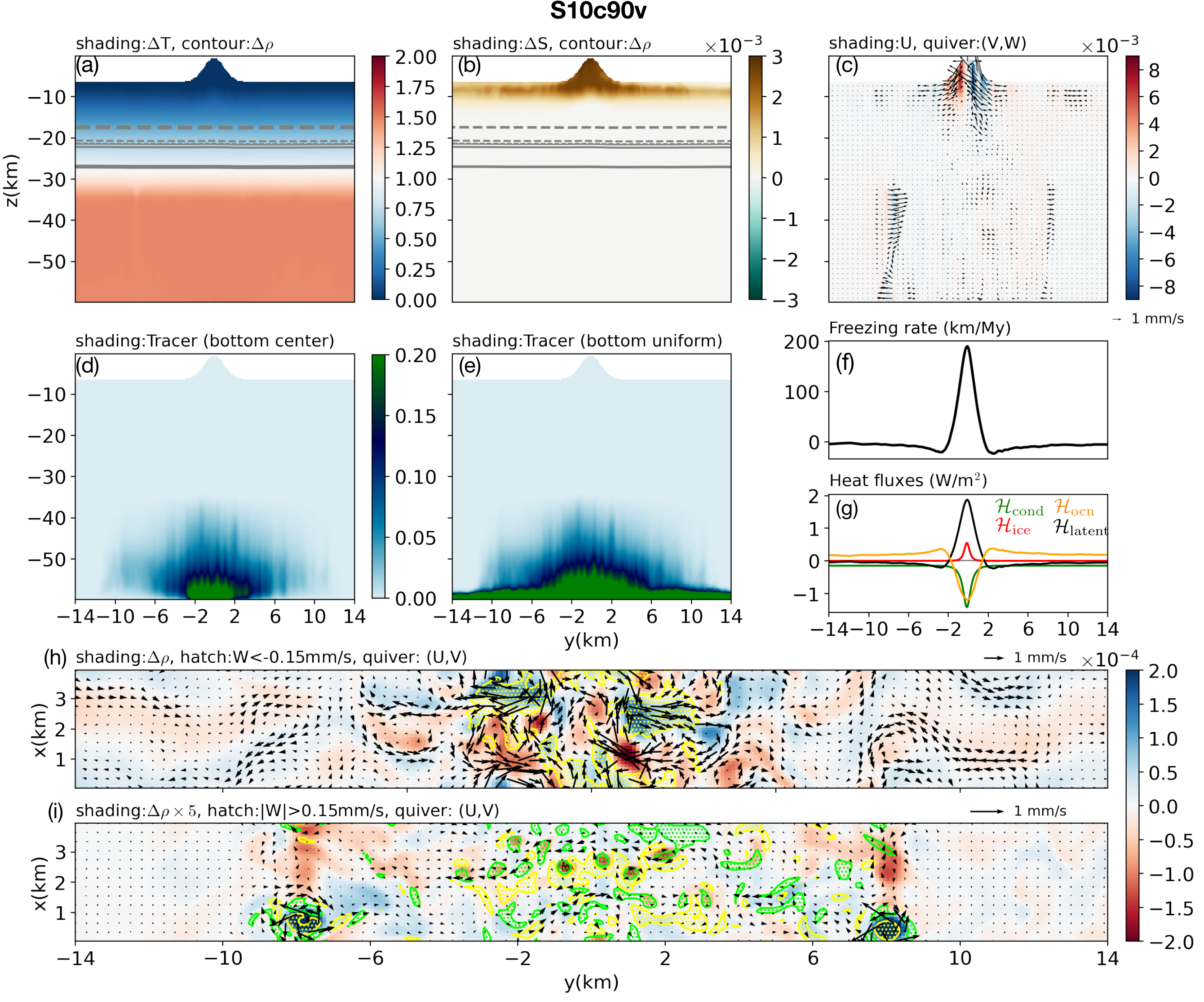}
    
    \caption{\small{Similar to Fig.~\ref{fig:S10c10}, except for the S10c90v scenario.}}
    
    \label{fig:S10c90v}
  \end{figure*}

  \begin{figure*}
    \centering
    \includegraphics[page=5,width=\textwidth]{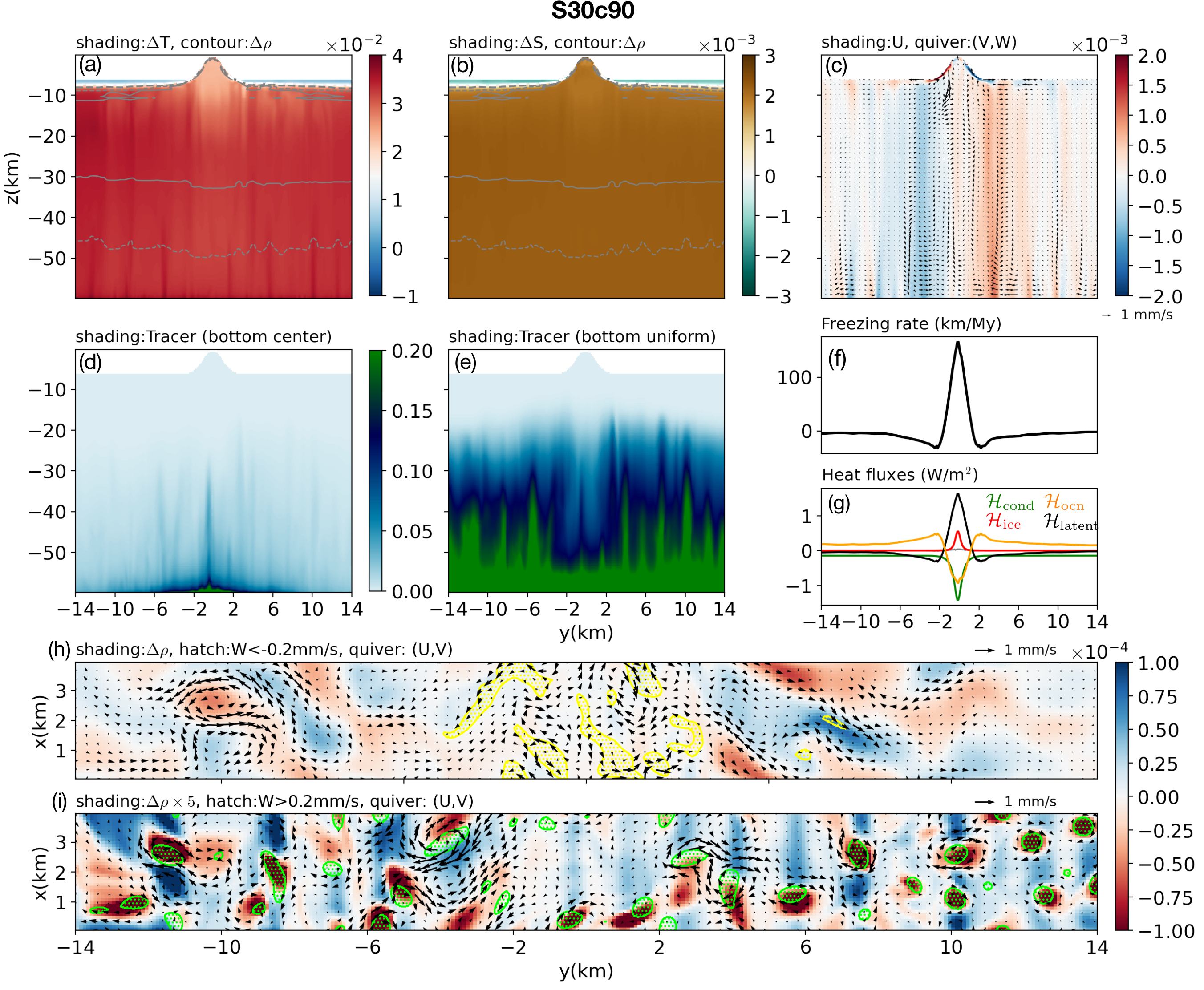}
    
    \caption{\small{Similar to Fig.~\ref{fig:S10c10}, except for the S30c90 scenario.}}
    
    \label{fig:S30c90}
  \end{figure*}

  \begin{figure*}
    \centering
    \includegraphics[page=6,width=\textwidth]{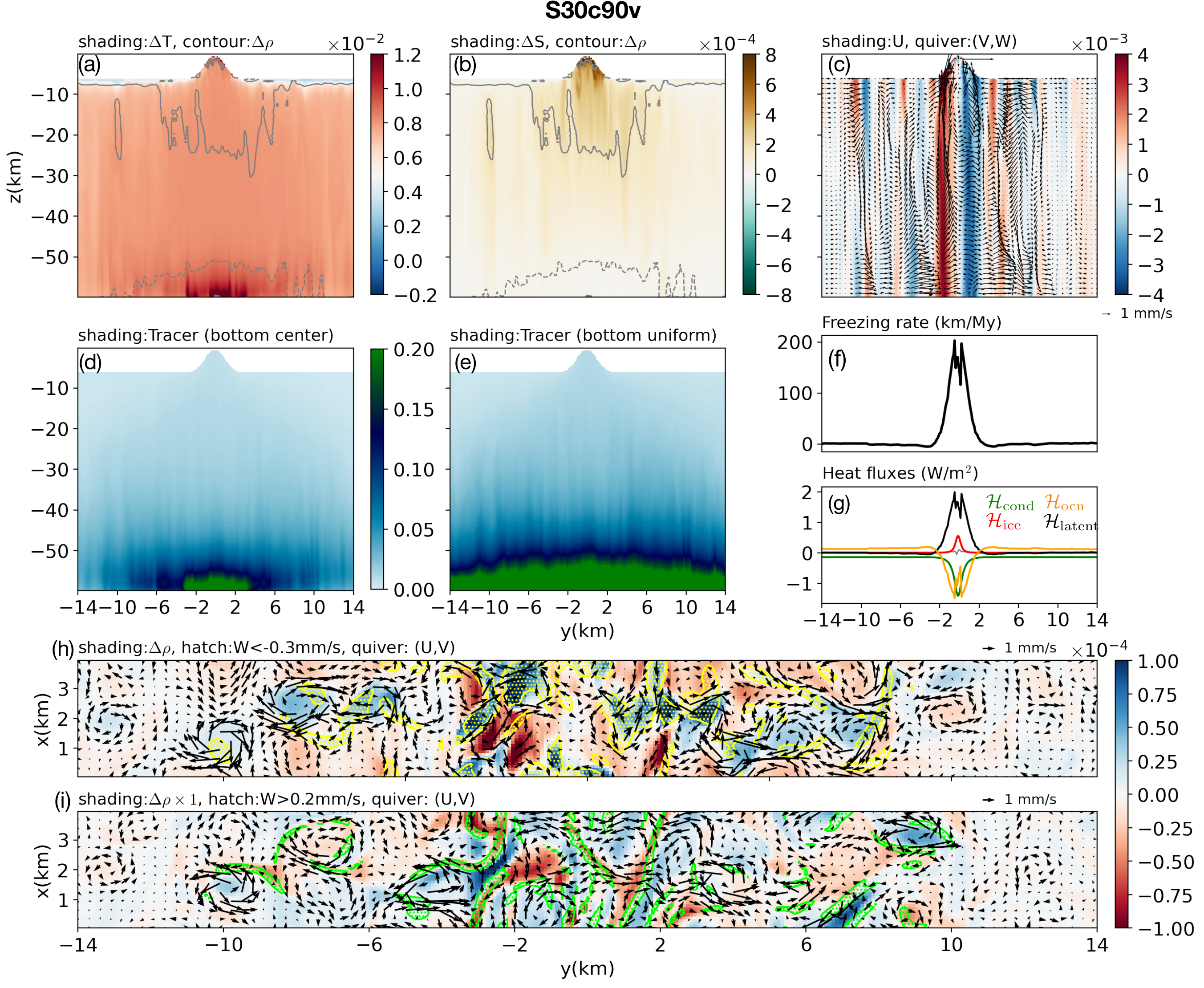}
    
    \caption{\small{Similar to Fig.~\ref{fig:S10c10}, except for the S30c90v scenario.}}
    
    \label{fig:S30c90v}
  \end{figure*}

    
    

    
    

   \begin{figure*}
    \centering
    \includegraphics[page=14,width=\textwidth]{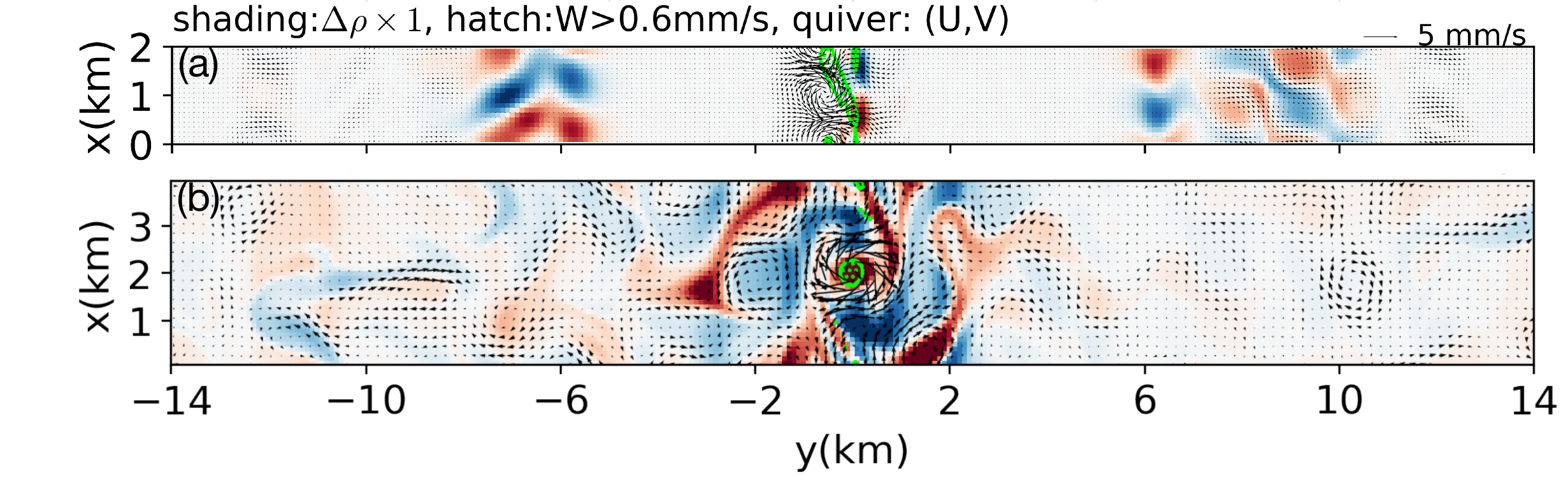}
    
    \caption{\small{Similar to Fig.~\ref{fig:S30c90v} panel (i). Panel (a) shows results for an experiment with half domain width, and panel (b) shows the initial stage of simulation after the domain is re-extended to the full width.}}
    
    \label{fig:S30c90v-narrow}
  \end{figure*}

  \begin{figure*}
    \centering
    \includegraphics[page=15,width=0.5\textwidth]{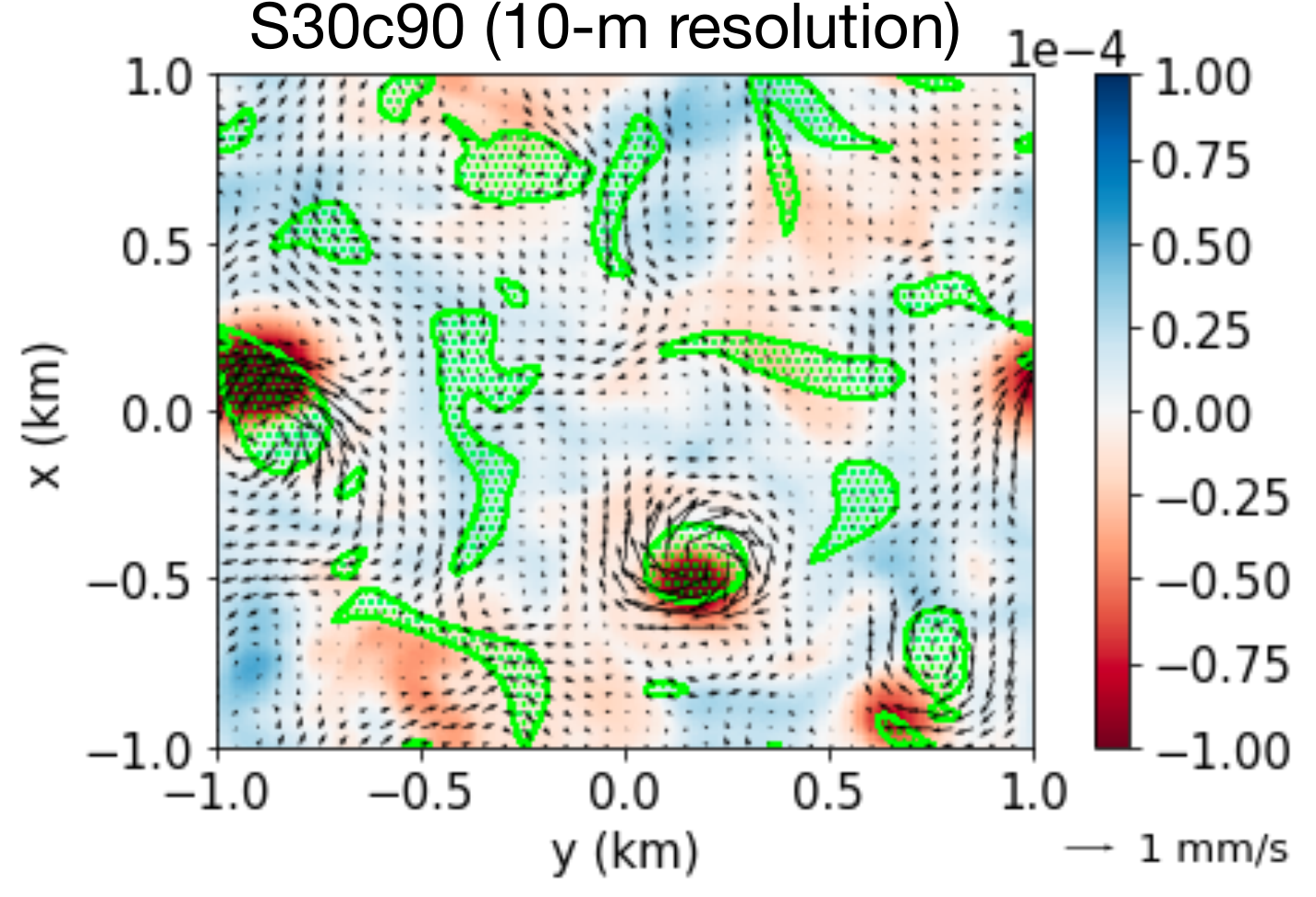}
    
    \caption{\small{Bottom ocean dynamics resolved by 10-meter resolution model for S30c90. The figure should be compared against Fig.~\ref{fig:S30c90}i or the Fig.4a in the main text. To keep the computational cost manageable, we reduce the domain size to only resolve the region far away from the geyser area.}}
    
    \label{fig:S30c90-10m}
  \end{figure*}


\bsp	
\label{lastpage}
\end{document}